\documentclass[letterpaper,twocolumn,10pt]{article}
\usepackage{usenix2024_SOUPS}

\usepackage{tikz}
\usepackage{amsmath}
\usepackage{algorithmic}
\usepackage{graphicx}
\usepackage{textcomp}
\usepackage{xcolor}
\usepackage{comment}
\usepackage{subfiles}
\usepackage{pgf-pie}
\usepackage{mdframed}
\usepackage{multirow}
\usepackage{lipsum}
\usepackage{comment}

\begin{document}

\date{}

\title{\Large \bf Evaluating Privacy Perceptions, Experience, and Behavior of Software Development Teams}

\def\plainauthor{Maxwell Prybylo, Sara Haghighi, Sai Teja Peddinti, Sepideh Ghanavati}

\author{
{\rm Maxwell Prybylo}\\
University of Maine
\and
{\rm Sara Haghighi}\\
University of Maine
 \and
{\rm Sai Teja Peddinti}\\
Google
\and
{\rm Sepideh Ghanavati}\\
University of Maine
} 

\maketitle
\thecopyright

\begin{abstract}
With the increase in the number of privacy regulations, small development teams are forced to make privacy decisions on their own. In this paper, we conduct a mixed-method survey study, including statistical and qualitative analysis, to evaluate the privacy perceptions, practices, and knowledge of members involved in various phases of the Software Development Life Cycle (SDLC). Our survey includes 362 participants from 23 countries, encompassing roles such as product managers, developers, and testers. Our results show diverse definitions of privacy across SDLC roles, emphasizing the need for a holistic privacy approach throughout SDLC. We find that software teams, regardless of their region, are less familiar with privacy concepts (such as anonymization), relying on self-teaching and forums. Most participants are more familiar with GDPR and HIPAA than other regulations, with multi-jurisdictional compliance being their primary concern. Our results advocate the need for role-dependent solutions to address the privacy challenges, and we highlight research directions and educational takeaways to help improve privacy-aware SDLC.
\end{abstract}


\section{Introduction} 
With the vast increase in privacy violations in the US and around the world~\cite{globalbreaches}, many countries have adopted new privacy regulations~\cite{unctad_2021}, such as the European General Data Protection Regulation (GDPR)~\cite{GDPR}. 
With these new regulations, developers are under increased scrutiny while implementing privacy engineering solutions throughout the Software Development Life Cycle (SDLC) or face financial penalties. Many mobile apps are initially developed by a small team of independent developers with limited privacy expertise or access to legal/policy resources to make privacy decisions~\cite{balebako2014improving,balebako2014privacy, Alomar2022}. Research shows that this lack of access to privacy expertise leads to challenges in creating concise, accurate and consistent privacy policies~\cite{zimmeck2019maps, slavin2016pvdetector, slavin2016toward, okoyomon2019ridiculousness, bhatia2016privacy, breaux2015detecting,yu2016can, rodriguez2023comparing, jain2023atlas}, implementing privacy concepts throughout the SDLC - from early analysis to testing~\cite{ekambaranathan2021money, hadar2018privacy, Tahaei2021Springer, Alomar2022}, and distinguishing between privacy and security approaches, tools and regulations~\cite{spiekermann2018inside,spiekermann2018understanding,balebako2014privacy,dalela2021mixed,tahaei2021developers,Green2016DevelopersAN,Bednar2019,hadar2018privacy}. 


In recent years, several approaches, including  Privacy by Design (PbD), have been introduced to help developers incorporate privacy rules throughout the SDLC~\cite{GAP07, GAR14, GAP09, zimmeck2019maps, okoyomon2019ridiculousness, cranor2009, cavoukian2009, Hoepman2014PrivacyDS, hoepman2018making, colesky2016critical, Jain2022, jain2023}. 
However, few works examined the implementation of these solutions from the developers' perspective and their impact on privacy practices. 
Most studies focus on only a limited group of developers and overlook the broader SDLC roles and the unique challenges faced by each role (e.g., product manager when defining privacy requirements or the QAs when identifying privacy leaks)~\cite{macleod2017documenting,iwaya2023privacy, dalela2021mixed, tahaei2020understanding, Horstmann2024ThoseTA}. They also do not examine how factors such as legal expertise,  regulations, and regional differences influence software teams' privacy perceptions and practices. 

In this paper, we conduct a large mixed-method survey study on Prolific with 189 participants located in the US and 173 participants located in 22 other countries (in total 362), who are involved in various roles in the SDLC -- including administrators (e.g., scrum masters, product managers), development and Quality Assurance (QA) teams, and information security/privacy experts. The non-US participants are located in EU+UK (132), South Africa (21), Mexico (15), Canada (3), and South America (2). 
Our goal is to identify the current state of privacy comprehension, practices, and behaviors in various SDLC roles, and the privacy gaps that have yet to be addressed. Our survey comprises of three parts: pre-screening questions (e.g., describing their product/customers), generic questions regarding participants' demographics (education, role, company size, etc.); and role-specific questions to examine their perceptions, experiences, and behaviors. 
We combine the participants' demographics (e.g., location data) with the role-specific responses to help determine: 

\noindent\textbf{-- RQ1:} Are there any differences in privacy perceptions among various roles, locations, and other demographics? 

\noindent\textbf{-- RQ2:} Does access to privacy experts (e.g., a Chief Privacy Officer - CPO) impact privacy perceptions and practices? 

\noindent\textbf{-- RQ3:} How do privacy practices and experiences vary according to SDLC roles, locations, and other demographics?  

\noindent\textbf{-- RQ4:} What is the degree of familiarity of different roles regarding privacy concepts, approaches, tools, and regulations? 

To the best of our knowledge, this is the first study to conduct such a holistic evaluation based on the roles in the SDLC.

Our results show that participants have diverse perceptions/definitions of privacy, showcasing the need for a refined approach to privacy in SDLC. Scrum masters, product managers, and information security/privacy experts define privacy more in terms of limited disclosure, while developers and QAs perceive privacy as control over personal information. 
Our study finds a lack of adoption of most PbD strategies and other privacy techniques, such as Privacy Enhancing Technologies (PETs) and Privacy Impact Assessment (PIA), 
in SDLC. 
Most QA members rely on legal/privacy experts to protect users' data, and they lack privacy knowledge and expertise. Members of software teams are generally self-taught regarding privacy concepts, and most are not familiar with regulations that exist in the US, such as the California Consumer Privacy Act (CCPA) \cite{CCPA} and the Children's Online Privacy Protection Rule (COPPA) \cite{coppa}. We also find that software teams face challenges in both understanding and adhering to privacy regulations, 
especially across multiple jurisdictions. These findings highlight the need for more privacy-focused education and training. Comparing regional-specific trends regarding the use of PETs, the creation of PIAs, or the presence of a CPO, we did not observe any differences among our participants, regardless of their location. This shows that privacy practices are primarily determined by the culture of the organization and are not influenced by various regulations across regions \cite{arizon2021understanding}. Our results highlight the challenges faced in various SDLC roles and advocate the need for role-dependent solutions to address them. Based on these findings, we outline research directions and educational takeaways to help improve privacy-aware SDLC.

\label{sec:introduction}

\section{Related Work} 
 

Understanding developers' privacy expertise and concerns has been explored in research through user studies with developers and analysis of developers' forums~\cite{macleod2017documenting,iwaya2023privacy, dalela2021mixed, tahaei2020understanding, Horstmann2024ThoseTA}. 
Tahaei et al.~\cite{tahaei2021privacy} and Horstmann et al.~\cite{Horstmann2024ThoseTA}  conducted interviews with developers and privacy experts and identified factors such as poor privacy culture, tensions between privacy and other business rules, lack of proper communication between privacy experts and developers, lack of standardized privacy tools, and mismatch between the technical expertise of developers and privacy experts that impact how developers implement privacy. They also emphasize the role of privacy champions to minimize such barriers. 
In 2014 (i.e., pre-GDPR and CCPA), Balebako et al.~\cite{balebako2014privacy} examined how app developers make privacy and security decisions and revealed that smaller companies exhibit fewer positive privacy and security behaviors. Their research emphasizes the need for simplified, cost-effective privacy tools such as privacy checklists, especially for small firms. Other studies~\cite{iwaya2023privacy, alomar2022developers, khandelwal2023unpacking, Horstmann2024ThoseTA} with practitioners and developers highlight that while regulations impact practitioners' behaviors and corporate cultures, the developers and practitioners mostly rely on app markets to spot privacy issues, and they struggle with implementing and maintaining privacy labels, as well as leveraging third-party tools to maintain compliance.   
 
The analysis of Stack Overflow (SO) posts shows that developers frequently query regarding PbD, compliance, and confidentiality ~\cite{tahaei2020understanding, tahaei2022privacy}. Delile et al.~\cite{delile2023evaluating} compared privacy questions on SO with responses generated by ChatGPT to identify whether ChatGPT could be used as an alternative tool. Their results indicate that, in $\sim$30\% of cases SO is more accurate than ChatGPT. Li et al.~\cite{li2021developers} and Parsons et al.~\cite {parsons2023understanding} studied posts on several Reddit forums and identified that most discussions on personal data usage occur in response to external events such as Android OS changes or privacy laws. 

These studies pinpoint developers' challenges in correctly implementing privacy requirements and maintaining compliance. Our work complements these efforts; however, it is the first study to assess privacy perceptions, practices, and knowledge of members of software teams involved in various roles in SDLC through a large-scale mixed-method approach. Prior work focused only on developers (i.e., programmers) in the US and a few countries, whereas we studied members from various SDLC roles (including product managers, QA, etc.) spanning 23 countries. In our work, we investigate how factors such as organizational aspects (e.g., the presence of a CPO) and participants' demographics (e.g., role, education, and location) impact privacy perceptions, experience, and behaviors of software teams. We also explore how frequently developers use online forums for privacy-related queries.

\label{sec:related-works}


\section{Study Design}
In this paper, we aim to understand how members of software teams in small, medium-sized, and large companies (i.e., with <20, 21--100, and 100+ employees), implement privacy in their software applications and examine their level of privacy comprehension, expertise, practices, and behaviors based on various demographics (such as roles, location, education level, etc.). For this purpose, we first conducted a pilot study to evaluate our survey design and then a large-scale study with members of software teams in 23 countries. Our pilot study was completed in January 2023, while our large-scale study was done between February--April 2023.

\subsection{Survey Tools}
\label{sec:survey-tool}

\paragraph{Survey Creation}
\label{sec:survey-creation}
We utilized \emph{Qualtrics} for survey creation, a platform supported by our university. 
Using Qualtrics, we customized our survey to individualize questions based on the participants' role (Q9) 
as defined in Table \ref{table:participant_roles}. For example, we asked developers about familiarity with PbD (Q39) and their use of forums such as Reddit (Q32), while information security members were asked about the management of access control, encryption algorithms, and certificates (Q60-Q62). 


\begin{table}[tb]
\centering
\caption{Breakdown of Participants Roles}
\begin{tabular}{|l|c|}
\hline
\textbf{Role} & \textbf{Count} \\ 
\hline
AD: Admin., Product Manager, Scrum Master & 70 \\ 
\hline
SD: Software Designer, Architect, Developer & 198 \\ 
\hline
QA: Software Tester, Quality Assurance Eng. & 40 \\ 
\hline
ISec: Information Security/Privacy Expert & 54 \\ 
\hline
\textbf{Total} & \textbf{362} \\
\hline
\end{tabular}
\label{table:participant_roles}
\end{table}

\noindent\textbf{Survey Platform Selection}
\label{sec:survey-platform}
We conducted the large-scale survey using Qualtrics integration on the Prolific~\cite{palan2018prolific, Tahaei-CHI2022} platform, since it provides a higher pay rate and allows selecting from a more specific pool of participants with basic programming knowledge, in our case - software teams. 

Tahaei et al. and Kaur et al.~\cite{Tahaei-CHI2022,kaur2022recruit} recommend using Prolific and MTurk for large-scale surveys.
Although pre-screening via programming questions is recommended~\cite{Tahaei-CHI2022, kaur2022recruit, danilova2021you, serafini2023recruitment}, it has limitations: (a) overusing such questions could lead to automatically responding \emph{correctly} without paying attention to the questions~\cite{Tahaei-CHI2022}; (b) in studies such as ours where the software teams include a variety of roles (e.g., product manager, QA, etc.) as well as with specific programming skills (e.g., JavaScript developers), having programming knowledge questions may bias the participants' pools towards more experienced developers in larger companies with traditional programming knowledge, preventing recruiting \emph{novice} developers and those in other SDLC roles; 
and (c) AI tools like ChatGPT~\cite{brown2020language} are widely accessible and can handle code-based questions. Thus, these questions are no longer a strong barrier to screen participants. Our analysis of Danilova et al.'s~\cite{danilova2021you} pre-screening questions with GPT-3.5 shows that the tool can answer the questions with 95\% accuracy. We discuss how we mitigate these issues below and in Section~\ref{sec:ethical-considerations}. 

\noindent\textbf{Conducting the Survey}
\label{sec:survey-conduct}
 Prolific maintains a pool of active participants who are regularly screened and vetted by the platform. In our survey, we decided on the sample size based on Prolific’s guidelines (a minimum of 300 for a representative sample). We initially pre-screened the Prolific participants based on the following requirements: (a) to be at least 18 years old, (b) fluent in English, and (c) working in industries such as Graphic Design, Information Services, Data Processing, Product Development, Software, Video Games, etc. We used their industry (rather than their role) as a filter, since Prolific does not allow selecting participants based on role. We paid an average of \$25.17/hr to those who completed the survey.  
 After the initial pre-screening, we recruited 686 participants across both US and non-US pools. Out of the 686 participants who started the survey, 14 did not give their consent, and 295 did not finish the survey; hence, they were excluded from our analysis. 
We then conducted another filtering process to ensure that the participants work in the software industry and, in fact, have software development experience. We asked them, \textsf{``Q4. In short, tell us about your product and who your customers are.''} We manually evaluated their responses and cross-checked them with Q6 (their post-secondary degree) and Q9 (their roles). We found that most of them are involved in software development activities such as \emph{``I make a productivity app for Mac \& Windows to record \& share the user's screen.''} We eliminated 15 participants as we could not verify their involvement in SDLC; for example, those with responses as \emph{``NA''} or \emph{``I sell home decor items. My customers are primarily women.''} Following these steps, we ended up with a total of 362 participants for our final count.
 

\subsection{Pilot Survey: University Students}
\label{sec:pilotstudy}

Our pilot survey participants were our university's graduate students (who mostly have industry experience through internships and part-/full-time jobs) over the age of 18 from the disciplines of Computing and Information Science, Electrical and Computer Engineering, and Business, who had experience in software development, IT, or related fields. 
To maintain their anonymity, we did not collect any personally identifiable information such as their contact, names, or company names. 

The goal of the pilot study was to gather initial insights and feedback before the deployment of our main study on Prolific. Upon the IRB approval, we launched the survey using Qualtrics. The survey consisted of 40 questions, including 13 short and 27 multiple-choice questions, which were derived based on our informal discussions with developers in small companies and prior gaps in research. We estimated that the survey takes $\sim$30-40 minutes to complete. Every participant was presented with the same set of questions regardless of their role on a software team. We used the responses to improve our large-scale survey  (i.e., Subsection \ref{sec:mainstudy}).

We received 45 responses but most were incomplete due to the survey's length and the diversity of questions. After discussing the study with the participants, 
we revised and shortened the survey based on participants' role in the SDLC.

\subsection{Software Teams Survey} 
\label{sec:mainstudy}

The main feedback we received from the pilot study was that the survey required too much time to complete ($\sim$27 minutes). To address this limitation and to focus on capturing participants' perspectives related to their SDLC roles, we separated the survey questions according to the roles. This shortened the survey duration by 12 minutes and enhanced the quality of the responses we received. We first asked all participants the same set of 10 questions that are partly related to demographics and the degree of privacy understanding. 
We then divided the remainder of the questions into four groups, one for each role defined in Table~\ref{table:participant_roles}.
Our breakdown loosely follows the SDLC phases, but we separated the Information Security/Privacy (ISec) roles from the Software Developer (SD) roles to evaluate the significance of security or privacy knowledge in our survey. Although ``Others'' role was an option, none of the participants selected it.  

\subsection{Survey Questions}
\label{sec:survey_questions}
The survey includes a mix of demographic, perception, experiential, and behavioral questions which are crafted based on our RQs (see Section~\ref{sec:introduction}) and the challenges identified in prior research regarding creating privacy-preserving applications, such as understanding privacy concepts~\cite{hadar2018privacy, Bednar2019, aljeraisy2021privacy}, knowledge of regulations and establishing compliance~\cite{Amaral2021, ezzini2021, GAR14}, creating consistent and accurate privacy policies~\cite{zimmeck2019maps, breaux2014eddy, liu2018large, slavin2016pvdetector, gorla2014checking, pandita2013whyper, qu2014autocog, rodriguez2023comparing, jain2023atlas,yu2016can}, knowledge of privacy approaches and existing tools~\cite{gustavsson2020assessment, Hadar2018, bu2020privacy, hadar2018privacy}.   
The complete list of questions (except questions 1-3, which are the required Prolific identification questions and our consent form) is found here.\footnote{Survey questions: http://tinyurl.com/2p9n49e4} 

\textit{Demographic questions}
collect basic information about the participants, such as age, education, their SDLC role, and the company size; e.g.,
``\textsf{What areas/roles of the development team are you currently involved with?}''. 


\textit{Perception questions}
 aim to understand participants' perceptions toward privacy; e.g., ``\textsf{How do you define privacy?}''. 

\textit{Experiential questions}
ask about their experience with privacy challenges and tasks; e.g., 
``\textsf{What was the process for the Privacy Impact Assessment, and who was involved?}''.

\textit{Behavioral questions}
ask about the participants' behaviors and knowledge related to privacy; e.g., ``\textsf{List any privacy-by-design strategies you have used or know.}''

\label{sec:survey-methods}

\section{Ethics \& Limitations}

\textbf{Ethical Considerations} This research adheres to our university's ethical guidelines and was conducted with our Institutional Review Board (IRB)'s approval. All participants agreed to a thorough consent form that included information about the investigators, the risks, benefits, compensation, and confidentiality. All participants were informed about their voluntary participation, maintaining their right to withdraw at any time. No personally identifiable information was collected, and measures were in place to ensure the anonymity, confidentiality, and security of responses. The contact information of all investigators and the IRB team was also included. No participants contacted the investigators or the IRB about the study or the compensation. 

\noindent\textbf{Limitations} Like most survey studies, our analysis is based on participant self-report data and is affected by self-report bias, recall bias, and social desirability bias. Participants were informed during consent that the survey pertained to privacy due to our institutions’ IRB requirement. 
This may introduce priming and self-selection biases.
There is also recruitment bias as the Prolific user base may not fully represent the diverse population of SDLC individuals. We used multiple screening questions to ensure that recruited participants have experience in software development activities (Section \ref{sec:survey-tool}). We adopted a conservative process to remove participants for whom we could not verify their SDLC involvement, however, we may have removed a few professionals. We also asked follow-up and write-in questions to ensure the multiple-choice questions were backed up with written facts. To mitigate the potential for survey responses being generated by AI tools like ChatGPT~\cite{brown2020language}, we minimized open-ended questions in favor of multiple-choice formats and carefully scrutinized the write-in responses to remove those that appeared AI-generated. Short responses with typos and errors suggested that our responses were not AI-generated. Despite our efforts, AI-generated responses could affect the study's outcomes.

We carefully framed our questions so as not to prompt biased responses. However, we could not avoid one leading question that asks about the confidence in their companies' privacy and security measures. We aimed to reduce the bias by providing four options instead of a `yes' and `no', with the option to not answer. Additionally, the question follows their own definition of privacy, further helping minimize bias. We employ statistical analyses (like the chi-square test~\cite{greenwood1996guide}) to ensure the broad applicability of our findings. To control for Type I errors in the presence of multiple hypothesis tests, we report our results after employing Bonferroni correction. 
\label{sec:ethical-considerations}

\section{Study Analysis Process}
Our survey results are organized around our research questions (RQs, see Section~\ref{sec:introduction}), focusing on various areas of privacy within the SDLC and across different roles. Our RQs examine the perceptions held, privacy experience and challenges, and privacy behaviors while 
considering the demographic breakdown (see Section \ref{sec:survey_questions}) to provide additional context and to allow for a more nuanced understanding of the data. Our analysis follows a mixed-method approach, encompassing both quantitative and qualitative methodologies. 

\textbf{Qualitative Analysis} We evaluate the descriptive and open-ended questions through open coding procedures and iterative processes. However, in our analysis, we used taxonomies and categories based on the current literature to classify the responses.  
For the open-ended question regarding the \emph{definition of privacy}, the first two authors, independently, classified 50 responses based on the taxonomy of privacy introduced by Solove~\cite{solove2006taxonomy} and the examples and hypotheses from~\cite{IAPP-Taxonomy, Harkous2022}. Similarly, for the \emph{usage of PETs}, we used PETs categories from the literature \cite{shen2011privacy, danezis2015privacy, PETsMatrix-ENISA}. The first two authors independently assigned categories for the first 25 responses. They then
discussed their results, resolved the discrepancies, and created a guideline (see Appendix~\ref{sec:appendix_qualitative}). They continued with the rest of the responses, evaluated the agreements and resolved the disagreements in another round of discussion. 
Lastly, a third privacy expert examined the results to ensure their correctness and completeness. For the non-subjective descriptive questions e.g., \emph{which PbD strategies they use}, one author categorized them based on the current literature, (e.g., privacy by design strategies~\cite{hopemanprivacy}, phases and roles in the SDLC~\cite{SDLC2010} for PIAs) and the second author reviewed them for correctness. 

\textbf{Quantitative Analysis} For the questions where our goal is to understand if a correlation exists between the demographics and the privacy-related perception, experience, and knowledge, we conducted statistical analyses. We used the Chi-Squared test~\cite{greenwood1996guide} 
to determine whether there is a significant correlation between two categorical variables. For questions where the responses are on a Likert scale, we used the Kruskal-Wallace test~\cite{breslow1970generalized}. For \emph{perception}, \emph{experience}, and \emph{behavioral} questions, we hypothesize from our RQs that the size of the company, the presence of a CPO or a similar role, the education level, roles, and participants' location may impact their confidence in privacy/security measures, various privacy practices (such as the creation of PIA or privacy policies), and their familiarity with PETs, regulations, and usage of forums. 
To control Type I errors and avoid false positives, we use Bonferroni correction \cite{Bonferroni}. Since Bonferroni correction is very conservative and may increase Type II errors, we discuss the results with respect to $alpha = 0.05$ as well as the adjusted value (i.e., $\frac{0.05}{24} = 0.0021$, for our 24 statistical tests). 

\label{sec:survey-results}

\section{Findings}
\subsection{Survey Demographics}
\label{sec:demographics}
 
In our main study, we received a total of 362 responses (after filtering - see Section~\ref{sec:survey-tool}). 189 participants reside in the US and the other 173 come from 22 other countries (see Section \ref{sec:introduction}).
Table \ref{table:participant_roles} shows a breakdown of participants' roles, with the majority ($\sim$55\%) in SD roles.
As shown in Appendix~\ref{sec:appendix_demographic} - Table \ref{table:demographics}, most participants identify as male, are below the age of 45, and have completed their BSc., with $\sim$61\% in Computer Science (CS), Information Technology (IT), Data Science (DS), and Electrical \& Computer Engineering (ECE) majors. This value includes the answers to ``Others, please specify''. Among those with a Business degree, 61\% are in AD (e.g., product manager), and 28\% are in SD roles. Among those in the ``Other'' degree category, 48\% identified as SD, 19.5\% as QA, 21.0\% as AD, and 11.5\% as ISec. The company sizes of $<$100 and 100+ employees are distributed almost equally. 

\subsection{Perceptions of Privacy}
\label{sec:perception-privacy}

We seek to understand software teams' privacy comprehension by examining how they define privacy, their confidence in their company's practices, and if these differ based on roles or organization differences (i.e., RQ1\&2).

\subsubsection{Definition of Privacy}
One of our key questions is, ``\textsf{How do you define privacy?}''. 
The responses were diverse, showing differing perceptions. Some participants defined privacy in terms of data security, highlighting the need to protect user data from unauthorized access. For example, one participant explained that \emph{``It involves implementing measures to safeguard sensitive information, such as encryption, access controls, and data anonymization"}. Others described privacy from a user rights perspective: 
\emph{``I define privacy as the ability to control all that is related to my information and to keep it from reaching someone who is unauthorized"}. Few responses incorporated legal compliance, with one participant defining privacy as: \emph{``This involves being compliant with regulations and ensuring all data is protected with a least-privilege access model with ownership of the different part data sources with assigned data stewards"}.

To categorize the diverse definitions of privacy, we utilized Solove’s taxonomy \cite{solove2006taxonomy}, that 
breaks down privacy into various categories based on the types of harm of a privacy breach. We chose Solove's taxonomy for two key reasons: (a) it provides a structured and detailed approach to understanding and analyzing definitions of privacy, which is essential with our wide range of definitions and perspectives; (b) it has been widely recognized and used in privacy research \cite{bamberger2011privacy,hoofnagle2010different,zimmer2008gaze,bennett2017governance}.  
We followed an open coding procedure to map the provided definitions with the taxonomy, as described in Section \ref{sec:survey-results}. 
Multiple classes for each definition were also possible. 
Figure \ref{fig:privacy_definitions} shows the mapping.  
(For a breakdown of Solove's Taxonomy see Appendix \ref{sec:appendix_solove} - Table~\ref{table:soloves_categories}; the `Blackmail' category did not apply to any participant's definitions.)

\begin{figure}[b]
    \centering
    \includegraphics[width=0.32\textwidth]{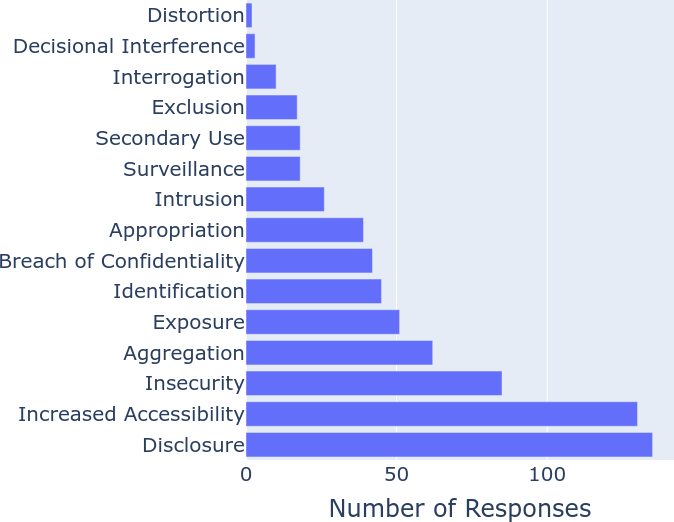}
    \caption{Privacy Definitions based on Solove's Taxonomy
    }
    \label{fig:privacy_definitions}
\end{figure}

Figure \ref{fig:privacy_definitions} shows that the top frequently occurring categories are `Disclosure', `Increased Accessibility', and `Insecurity'. 
This result indicates that most participants either consider the traditional definition of privacy as \emph{control over personal information} or perceive privacy in terms of \emph{security}. 
For `Disclosure', one participant highlights the importance of transparency and clear communication about data collection purposes and user control: 
\emph{``Privacy is the assurance that all data belonging to an individual will be disclosed to others only with that individual's consent, for uses understood and approved by that individual.''}
For `Increased Accessibility', a participant who works with genetic data underscores the need for controlled access to such information, only granting access if needed: 
\emph{``The users' ability to define who can access their data and even in that what kind of data can be accessed. As I work in genetic data from patients in my line of work, the clinical information is always controlled access and only researchers working on the particular project can gain access on a need-to-know basis.''}

We further examined how privacy perceptions differ across various roles. Almost 50\% of participants in AD or ISec roles define privacy as `Disclosure', while QA and SD roles mostly consider privacy as `Increased Accessibility', which is related to \emph{access control}. ISec roles mentioned `Aggregation' more frequently than other roles, which is an anonymization technique used only in privacy rather than security.


\begin{mdframed}[backgroundcolor=darkgray!20,linewidth=0pt,innerleftmargin=10pt,innerrightmargin=10pt,skipabove=5pt]
The variety in our participants' definitions of privacy shows the complexity of privacy perceptions, and the need for a holistic approach that covers a variety of aspects of privacy throughout the SDLC.   
\end{mdframed}


\subsubsection{Confidence in Security and Privacy Measures}
\label{sec:confidence}

We asked participants about their confidence in the privacy and security measures implemented in their organization.
Table~\ref{table:confidence} shows the distribution of the participants and their responses. Note that `PnS' stands for `prefer not to say'. In all roles, most participants are confident in their company's security and privacy measures. Interestingly, ISec members are the most confident while the QA members are the most uncertain. This can be either due to QA members considering privacy and security as an afterthought~\cite{hadar2018privacy}, thus ignoring these requirements, or because they encounter more non-compliance instances during testing than any other roles.


\begin{table}[t]
\centering
\caption{Distribution of Participants' Confidence}
\begin{tabular}{|l|c|c|c|c|}
\hline
\textbf{Role} & \textbf{Yes} & \textbf{No} & \textbf{Unsure} & \textbf{PnS} \\ 
\hline
\textbf{AD} & 53 (75.7\%) & 6 (8.6\%) & 11 (15.7\%) & 0 \\ 
\hline
\textbf{SD} & 149 (75.3\%) & 8 (4\%) & 35 (17.7\%) & 6 (3\%) \\ 
\hline
\textbf{QA} & 25 (63\%) & 2 (5\%) & 12 (30\%) & 1 (2\%) \\ 
\hline
\textbf{ISec} & 44 (81.5\%) & 2 (3.7\%) & 8 (14.8\%) & 0 \\ 
\hline
\textbf{Total} & \textbf{271} & \textbf{18} & \textbf{66} & \textbf{7} \\ 
\hline
\end{tabular}
\label{table:confidence}
\end{table}

We analyzed whether there is a correlation between participants' confidence in security and privacy measures and their demographic factors, such as the company's size (\textbf{H1a}), participants' roles (\textbf{H1b}), education level (\textbf{H1c}), and the presence of CPO or a similar position (\textbf{H1d}) (see Appendix \ref{sec:appendix-confidece} and Table \ref{table:hypothesis1} for more details). 
As shown in Table \ref{table:hypothesis1}, with Bonferroni adjustment ($\frac{0.05}{24} = 0.0021$), we cannot reject the null hypothesis for \textbf{H1a}, \textbf{H1b}, and \textbf{H1c} ($p-value= 0.494, 0.654$ and $0.570$); thus, we find no correlation between confidence in security and privacy measures and a company's size, participants' roles, or education levels. However, with a $p-value= 0.0007$ for \textbf{H1d}, we can reject the null hypothesis and say there is a correlation between the presence of a CPO (or similar position) and confidence in privacy and security measures. We further evaluate whether the existence of a CPO could lead to positive privacy outcomes in Subsection~\ref{sec:CPO}.

We asked the ISec members specific questions regarding their company's security and privacy measures. When asked \emph{``whether their company conducts security audits for third-party software used in their products''}, slightly more than half ($\sim$56\%) said `Yes' while a large number ($\sim$38\%) were `Unsure'. This is alarming since research shows a large number of third-party software and libraries include security and privacy vulnerabilities \cite{Zhan2021, HE2019259, alfadel2023empirical}. However, when we asked \emph{``whether their company securely manages encryption keys and implements encryption algorithms and access control policies''}, more than 70\% responded `Yes' -- which highlights inconsistencies in privacy practices even among experts.

\begin{mdframed}[backgroundcolor=darkgray!20,linewidth=0pt,innerleftmargin=10pt,innerrightmargin=10pt,skipabove=5pt]
A CPO is important in fostering employees' confidence in the privacy and security measures of an organization.
\end{mdframed}

\subsubsection{Presence of a Chief Privacy Officer (CPO)}
\label{sec:CPO}

To evaluate the impact of a CPO or other similar roles on privacy practices, we focus on the AD and SD roles, who are the majority of our participants (i.e., 268 (74\%)). We did not include ISec and QA teams to avoid any response bias, due to their active privacy role in the company. We asked them ``\textsf{Do you have a Privacy Officer or similar position in your company?}''.  
Table~\ref{table:cpo} in Appendix \ref{sec:appendix-cpo} shows the distribution. Interestingly, only slightly more participants responded `Yes' (42.6\%) than `No' (38.4\%). $\sim$18\% were `Unsure', which may indicate the lack of proper communication among employees regarding the company's privacy practices and the purpose of a CPO. 1.1\% responded `Other', which included a legal team or a CTO. Among those that said `Unsure', 23\% are in AD and 77\% are in SD roles which may indicate CPO members communicate more with the management team (i.e., AD). 

\begin{table}[bt]
\centering
\caption{Distribution of Company Size vs Existence of a CPO}
\begin{tabular}{|l|c|c|c|c|}
\hline
\textbf{Company Size} & \textbf{Yes} & \textbf{No} & \textbf{Unsure} & \textbf{Others} \\ 
\hline
\textbf{0--20}& 31.5\% & 51.5\% & 15.7\% & 1.4\% \\ 
\hline
\textbf{21--100}& 46.1\% & 29.2\% & 21.6\% & 3.1\% \\ 
\hline
\textbf{100+}& 47.3\% & 34.9\% & 17.8\% & 0\% \\ 
\hline
\end{tabular}
\label{table:company-cpo}
\end{table}

We investigated whether the larger companies have a CPO. Table \ref{table:company-cpo} shows the distribution of the presence of a CPO based on the company size. Here, we see the presence of a CPO increase with the company size. We also observe that companies of all sizes have a sizable number of `Unsure' responses.

We further asked the participants ``\textsf{When you have a question about compliance with regulations, what do you do?}''. The participants could select more than one option. Appendix~\ref{sec:appendix-cpo} - Table~\ref{table:compliance-questions} shows the distribution of the responses. About half of the respondents (50.1\%) mention they ask lawyers or a CPO, while 23.1\% look at the best practices and standards (such as NIST guidelines), and 18.5\% use developers' forums (such as Stack Overflow). Among `Other' sources, they mainly mention `search Internet' or `ask a colleague'. 

We analyzed whether the existence of a CPO (i.e., access to a legal or privacy expert) could impact the creation of PIA (\textbf{H2a}), the familiarity with PETs (\textbf{H2b}), 
the number of privacy breaches (\textbf{H2c}), or is influenced by the company size (\textbf{H2d}). 
Appendix \ref{sec:appendix-cpo} and Table \ref{table:hypothesis2} show the list of the hypotheses and the results of the tests.
With Bonferroni correction, our results show that the presence of a CPO correlates with the size of a company ($p-value <0.00001$). This correlation indicates that larger companies are more likely to have a CPO or a legal/privacy expert to help mitigate privacy risks, which is aligned with findings in \cite{balebako2014privacy}. However, with the p-value adjustment, we do not find a correlation between the presence of a CPO (or a similar position) and the creation of a PIA ($p-value = 0.1005$) and the use of PETs ($p-value = 0.008$). This may not be surprising, especially since a majority of SD roles, who are the main users of PETs and are involved in the PIA creation, are unaware of a CPO role.
Our analysis also did not reveal a significant correlation between the presence of a CPO and the number of privacy breaches experienced by the organization ($p-value = 0.359$). This suggests that the presence of a CPO, while important and necessary, may not be sufficient to help minimize privacy breaches. 

\begin{mdframed}[backgroundcolor=darkgray!20,linewidth=0pt,innerleftmargin=10pt,innerrightmargin=10pt,skipabove=5pt]
Although a CPO could improve confidence in a company's privacy measures, it has limited effectiveness
in enhancing privacy practices and reducing breaches. 
\end{mdframed}

\subsection{Experience with Privacy} 
\label{sec:experience}

We ask members of software teams in various roles about their \emph{experience} with creating privacy policies and/or PIA, as well as practices to ensure the protection of users' data to better understand their privacy challenges (i.e., RQ2\&3).

\subsubsection{Creation of a Privacy Impact Assessment}

A Privacy Impact Assessment (PIA) is a critical tool for identifying and mitigating privacy risks at any stage of software development. Recently, PIAs and their variations, such as the Data Protection Impact Assessments (DPIAs), have become a requirement in GDPR~\cite{GDPR} and CCPA~\cite{CCPA}. This tool allows organizations to address privacy and security issues before they become problems. 
In our survey, we asked participants if a PIA was created at any time throughout development, and if the answer is yes: at what stage it was created, who was involved, and what the process for creation was. 



We received 311 responses, where only 43 (14\%) of them (where more than half were outside of EU+UK) reported that they created a PIA at any point in the SDLC, while a significant proportion (57.2\%) reported that they did not (see Appendix \ref{sec:appendix-PIA} - Figure \ref{fig:creation_PIA}). This indicates a lack of awareness regarding the existence or the need for PIAs (i.e., the PIAs are non-existent or are conducted without their knowledge by the CPO or other teams). We also observed that $\sim$25\% are unsure about whether a PIA was created, which may highlight a gap in communication within a company about its privacy practices. $\sim$4\% chose `Prefer not to say'.

\begin{figure}[tb]
    \centering
    \includegraphics[width=0.35\textwidth]{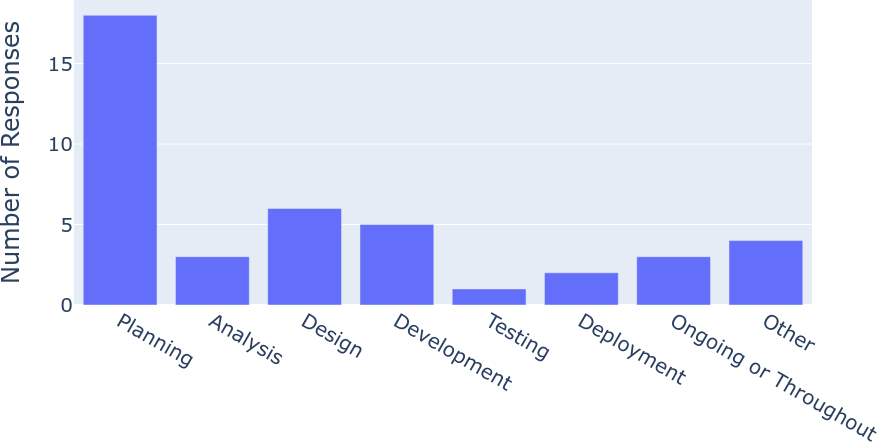}
    \caption{Stages of the SDLC When PIAs are Created. 
    }
    \label{fig:pia_sdlc_stages}
\end{figure}

Among the 43 who created a PIA, 3 ($\sim$7\%) did not answer the follow-up questions. Our results show that PIAs were created at various stages in the SDLC (see Figure \ref{fig:pia_sdlc_stages}), but $\sim$51.0\% are at the planning and analysis stages. One participant who reported that a PIA was created during the planning stage said, \emph{``At the start of development of idea because privacy is more important than all things''}. Some participants reported creating a PIA at the start of development, e.g., \emph{``We created a [PIA] at the beginning of the software development process. This allowed us to identify potential privacy risks and develop strategies to mitigate them''}. Others mentioned during the design, or even towards the end of development. 

The sizable number of participants (42\%) involved in PIA during the later stages in SDLC may indicate that privacy requirements are not considered early on, and are only included as an afterthought -- which is aligned with the findings in~\cite{hadar2018privacy}. Furthermore, the variation in the timing of the PIA creation shows the need for a more standardized approach to incorporating privacy considerations into software development. 

\begin{figure}[t]
    \centering
    \includegraphics[width=0.3\textwidth]{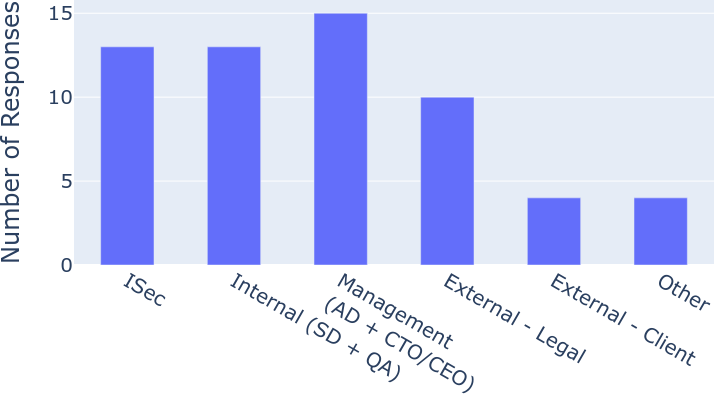}
    \caption{Distribution of Roles Involved in PIA Creation.
    }
    \label{fig:PIAProcess}
\end{figure}

Figure~\ref{fig:PIAProcess} shows the distribution of the roles involved in PIA creation. More than one category was allowed. The responses are also diverse. Some participants reported that they created the PIA themselves or it was a team effort (i.e, SD \& QA teams), while others reported that it was done by the CPO or external Legal team, ISec teams, upper management (i.e., CEO or CTO), or even the client (External). This shows that the responsibility for privacy can be distributed across various roles, which again highlights the need for clear communication, collaboration, and defined privacy practice processes. 

The responses to the question regarding the process for creating the PIA also varied. Some initiate the process by downloading a template and collaborating with internal specialists, while others seek guidance from lawyers, executives, or third-party experts. A common approach involved consulting with professionals, with one participant mentioning that they \emph{"outsourced [a] developer that specialises in data privacy and security"}. Several participants mentioned the involvement of specific roles, such as the Data Chief, IT teams, and privacy protection specialists. 
The process often involved cross-functional teams.
In some cases, senior leadership, e.g., the CTO, CEO, or owner, played a pivotal role in the process.

Lastly, we evaluated the PIA correlation between the company size (\textbf{H3a}) and the participants' confidence in privacy and security measures (\textbf{H3b}) (see Appendix \ref{sec:appendix-PIA}). 
The results show a significant correlation $p-value < 0.0001$ for both tests - even after Bonferroni correction.


\begin{mdframed}[backgroundcolor=darkgray!20,linewidth=0pt,innerleftmargin=10pt,innerrightmargin=10pt,skipabove=5pt]
Most members of software teams are not familiar with PIA or are unaware of its creation. However, those involved in PIA emphasize the need for its creation in the initial phases of the SDLC in a collaborative process with experts from various departments and consultants. 
\end{mdframed}

\subsubsection{Creation of Privacy Policies}

Privacy policies describe how, why, and how long an application uses personal information. Regulations~\cite{GDPR,CCPA} and the FTC~\cite{FTC} require companies to provide users with detailed privacy policies. Research shows that these policies may be inconsistent with apps~\cite{zimmeck2019maps, slavin2016toward}, since they are either created by outside legal experts (who may not fully comprehend the apps) or by using privacy policy generators~\cite{zimmeck2021privacyflash}. 

We asked the AD team about their experience and challenges with privacy policies (as other roles are often only indirectly involved). Out of the 70 participants, 3 did not provide any answer. Of the rest, only 11 (17\%) have been involved in the creation of a privacy policy, and they used `legal experts' the most (64.0\%), followed by `templates' (45.5\%), and `privacy policy generators' (36.4\%). More than one response could be selected. Two of them mentioned that they either `search the Internet' or `ask for team input', in addition to using privacy policy generators and templates. In 60\% (out of 45.5\%) of cases that used `templates', and in 50\% (out of 36.4\%) of cases that used `privacy policy generators', a `legal expert' has also been selected. 
This result matches with prior research that legal experts in a company are mainly involved in the privacy policy creation, which may lead to inconsistencies between the app and the policy~\cite{slavin2016toward}. 
We also noticed that those who said `Yes' are mostly from companies with less than 100 employees ($\sim$64\%) and with a CPO ($\sim$55\%). 

Finally, we asked the 11 participants who responded `Yes', ``\textsf{What challenges did you face when creating your privacy policy?}''. We received 10 responses. Six of them describe the challenges regarding compliance with regulations in multiple international jurisdictions, and understanding legal jargon, rules, and standards. One specifically had concerns regarding compliance, since they use privacy policy generators: 
\emph{``...differences between different countries and their requirements since we are international.''}
Five respondents describe their main challenge as ensuring completeness (i.e., covering all personal information), soundness, and language of privacy policies. E.g.: \emph{``Whether the wording I chose was going to cover all the bases I needed it to and whether it was clear and easy to understand.''} or \emph{``Which rules and text to inform users;...''}
One of those five respondents was also concerned about which template to choose. Four others did not find the process challenging since they trusted the legal expert to help. 

\begin{mdframed}[backgroundcolor=darkgray!20,linewidth=0pt,innerleftmargin=10pt,innerrightmargin=10pt,skipabove=5pt]
Compliance with regulations, and ensuring completeness and correctness are among the most common challenges in creating a privacy policy. Software teams use several tools besides legal experts to help create privacy policies. 
\end{mdframed} 

\subsubsection{Privacy Practices to Protect Users' Data}

We tailored some of the privacy practice questions based on the role, specifically for ISec, SD, and QA teams. 
We asked ISec members: \textsf{``How do you ensure that data collected from users is used only for intended purposes?''}. 
They discussed various approaches. $\sim$32\% emphasized the importance of documentation to ensure transparency and accountability, with one noting \emph{``the meticulous documentation of every step in the data usage process''}. Encryption emerged as a common theme, with participants mentioning sending encrypted documents and ensuring data is stored securely. A respondent states ``\textit{I would send documents encrypted and compressed into a zip file, and instruct them to delete the file once the information is accessed.}'' 
Limiting access to data is another frequent approach, with 30.2\% stressing the importance of restricting data access to only those who need it and maintaining logs to track any access. 16.98\% stated the significance of transparency, ensuring they only collect necessary data, obtaining user consent, and regularly monitoring data usage. A few (9.43\%) pointed out the importance of adhering to specific regulations, such as the Health Insurance Portability and Accountability Act (HIPAA)~\cite{hipaa} and the Family Educational Rights and Privacy Act (FERPA)~\cite{ferpa}. 16.98\% admitted to not having direct control over data but trusted their organization's protocols and training to handle data responsibly.

We also asked the same group \textsf{``How do you manage access to sensitive user data in your organization?''}. Role-based access controls, multi-factor authentication, and encryption are common strategies employed to safeguard sensitive information. One respondent shared, ``\textit{We limit access to systems based on who really needs to access that data.}''. Such measures ensure that only authorized personnel can access sensitive data, thereby minimizing potential breaches. Regarding data retention practices, only 47.17\% of respondents state that they have been involved in removing user data either after its predetermined lifespan or upon user request.

We asked the SD members:  \textsf{``If you encounter a privacy concern at any point in the software development process, what steps would you take?''}. 
More than 95\% of them take the concerns very seriously. For example, one participant mentions \emph{``run a risk assessment''} and another mentions \emph{``We take the app offline and start iteratively testing parts of the app to see where the privacy concern is.''} About 36\% deal with the concern internally to fix it and communicate it with the client and upper management. Another 35\% directly escalate it to their supervisors, while 20\% seek help from the ISec team or lawyers. A handful contact the client first. 

We asked the QA team: 
\textsf{``How do you verify that third-party systems used in your products are privacy compliant?''}. Similarly, we received diverse responses. Only 56.6\% confirmed that their companies conduct security audits of these third-party systems. Some mentioned the significance of conducting vulnerability assessments and penetration testing to ensure third-party systems' compliance (23\%). 
Some respondents discussed that they rely on reading privacy policies and contracts of third-party systems (28\%), while others emphasized the importance of legal agreements and monitoring data transfers (15\%). About 22\% admitted to not being directly involved in this process, placing trust in their organization's legal and security teams, which is aligned with findings in \cite{cobigo2020protecting}.

Lastly, regarding the QA teams' practices for testing for privacy breaches and data leaks, they emphasized the importance of understanding the data they work with and always being vigilant about potential breaches. Regular manual or automated testing is a common theme. $\sim$27\% of them mentioned the use of penetration testing, both internally and via third-party services. Others stressed the importance of using fake data during testing phases and ensuring that real user data is always encrypted and protected. $\sim$16\% of respondents admitted to not being directly involved but trusted their organization's protocols and cybersecurity measures.

\begin{mdframed}[backgroundcolor=darkgray!20,linewidth=0pt,innerleftmargin=10pt,innerrightmargin=10pt,skipabove=5pt]
The most common privacy practices among SD, ISec, or QA teams are documentation, auditing, and security techniques (such as access control and encryption). QA teams rely heavily on legal and ISec teams to ensure data protection and are less involved themselves.
\end{mdframed}

\subsection{Privacy Awareness and Behaviors}
\label{sec:awareness}

We assess privacy \emph{behaviors} based on familiarity with regulations, PbD, PETs, and such knowledge sources (i.e., RQ4).

\subsubsection{Familiarity with Privacy Regulations}

\begin{figure}[bt]
    \centering
    \includegraphics[width=0.42\textwidth]{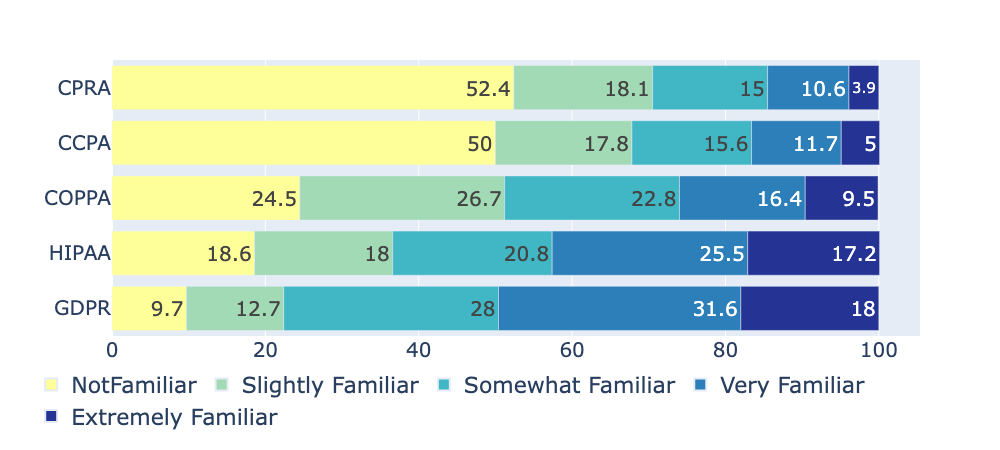}
    \caption{Familiarity with Different Regulations.
    }
    \label{fig:RegulationFig}
\end{figure}

In recent years, several regulations have been introduced that developers need to comply with. Non-compliance with these regulations may lead to financial penalties, sometimes up to 4\% of the annual turnover of the company~\cite{GDPR}. However, these regulations include legal terminologies that may not be familiar to members of the software teams. To understand the degree of familiarity and awareness, we asked all 362 participants about their familiarity with GDPR, HIPAA, COPPA, CCPA, and the California Privacy Rights Act (CPRA). The answers are on a Likert Scale (see Figure \ref{fig:RegulationFig}).

We combined the results from `somewhat familiar', `very familiar', and `extremely familiar' together and found that software teams' members, regardless of their region and roles, are more familiar with GDPR (77.35\%) and HIPAA (63.26\%). COPPA, CCPA, and CPRA are all 50\% or below. 
ISec teams are the most familiar with all regulations among all roles, followed by the SD and AD teams. 
The QA teams are the least familiar with 7.5\% familiarity with CCPA and CPRA, and 65.0\%, and 57.5\% with GDPR and HIPAA. 

We asked participants \textsf{``How did you learn about the previous regulations?''}. 
More than one option could be selected. As shown in Table \ref{table:regulations-learn}, the majority are self-taught while university education ranks second. Among all roles, the ISec team has the highest percentage of learning about regulations through university education (33.3\%), which is more likely through cybersecurity courses. We also asked the participants to describe the other sources they used to learn about privacy regulations. In most cases, they mentioned `training at work' as the source; however, 2 participants mentioned `social media' and `YouTube' as their source. 
\begin{mdframed}[backgroundcolor=darkgray!20,linewidth=0pt,innerleftmargin=10pt,innerrightmargin=10pt,skipabove=5pt]
GDPR is the most familiar regulation among all participants due to its comprehensiveness. ISec teams are more likely to learn about regulations through university education; hence, are more familiar with them than other groups. QA teams are the least familiar. 
\end{mdframed}

\begin{table}[t]
\centering
\caption{Distribution of Participants' Learning Experience}
\begin{tabular}{|l|p{0.9cm}|p{0.9cm}|p{1.4cm}|p{1.0cm}|p{0.9cm}|}
\hline
\textbf{Role} & \small\textbf{Self Taught} & \small\textbf{Lawyer} & \small\textbf{University Education} & \small\textbf{IAPP Cert.} & \small\textbf{Others} \\ 
\hline
\textbf{AD} & 57.1\% & 5.7\% & 20.0\% & 1.4\% &  15.8\% \\ 
\hline
\textbf{SD} & 56.6\% &3.5\% & 21.7\% & 2.5\% &  15.7\%\\ 
\hline
\textbf{QA} & 85.0\% & 7.5\% & 2.5\% & 0.0\% &  5.0\% \\ 
\hline
\textbf{ISec} & 42.6\% & 3.7\% & 33.3\% & 9.3\% &  11.1\%\\ 
\hline
\textbf{Total} & \textbf{57.7\%} & \textbf{4.4\%} & \textbf{21.0\%} & \textbf{3.1\%} & \textbf{13.8\%}\\
\hline
\end{tabular}
\label{table:regulations-learn}
\end{table}

\begin{table*}[th]
\centering
\caption{Distribution of PbD Strategies Used by Developers}
\begin{tabular}{|c|c|c|c|c|c|c|c|}
\hline
\textbf{Minimize} & \textbf{Hide} & \textbf{Separate} & \textbf{Abstract} & \textbf{Inform} & \textbf{Control} & \textbf{Enforce} & \textbf{Demonstrate} \\ 
\hline
21 & 22 & 7 & 2 & 17 & 12 & 1 & 4 \\ 
\hline
\end{tabular}
\label{table:PbDDistribution}
\end{table*}

\subsubsection{Familiarity with Privacy by Design (PbD)}

Privacy by design (PbD) strategies introduced by Hoepman et al.~\cite{hoepman2018making} have gained interest in helping developers to be compliant with regulations. 
We asked the SD members (i.e., 198 participants) if they are aware of PbD, and if they answered yes, whether they used them  (see Appendix~\ref{sec:PbD-Appendix} - Table~\ref{table:PbDFamiliairty}) and to list the ones they used. $\sim$46\% are familiar with PbD approaches while $\sim$25\% are unsure, which indicates the potential knowledge gap and opportunity for educating developers. Out of those who answered `Yes' to the awareness of the PbD question, only 57.1\% had employed such strategies in their work. Of the remaining, 23.1\% did not use them and 16.5\% were unsure. This result suggests that even among developers who are familiar with such strategies, not everyone acts on this awareness -- which may indicate the lack of usability and readiness of PbD for day-to-day developers' tasks \cite{tahaei2019survey} or other organizational factors, such as lack of resources \cite{iwaya2023privacy}.

Lastly, we evaluated the responses about the usage of specific PbD strategies (multiple answers were possible). Interestingly, our results are aligned with the findings of Tahaei et al.~\cite{TahaeiLiVaniea+2022+114+131} (see Table~\ref{table:PbDDistribution}). Our top categories are `hide' (22), `minimize' (21), `inform' (17), and `control' (12), while `enforce' (1) and `abstract' (2) are rarely discussed. One participant mentions \emph{``Mostly minimise. Its the most straightforward.''} This response reinforces our result in that `minimize' is one of the easiest strategies to implement. We also received responses regarding Anne Cavoukian's PbD principles~\cite{cavoukian2009} such as `privacy by default' (6 times) and `proactive' (twice). The use of PIA was also mentioned 5 times as a strategy. 
\begin{mdframed}[backgroundcolor=darkgray!20,linewidth=0pt,innerleftmargin=10pt,innerrightmargin=10pt,skipabove=5pt]
Our findings show that PbD approaches are not yet commonly used, and their lack of adoption underscores the gap in developers' knowledge regarding PbD and their usability in day-to-day developers' tasks. 
\end{mdframed}

\begin{table}[H]
\centering
\caption{Usage of PETs in Software Development Process}
\begin{tabular}{|p{6.1cm}|p{1.4cm}|}
\hline
\textbf{Privacy Enhancing Technology (PET)} & \textbf{Percentage} \\
\hline
Encryption & 70.48\% \\
\hline
Access Control/Identity Protection & 34.29\% \\
\hline
Anonymity and Pseudonymity & 9.52\% \\
\hline
Differential Privacy Approaches & 8.57\% \\
\hline
Secure Communication/VPN & 8.57\% \\
\hline
Privacy-Enhanced Anti Web Tracking & 0.0\% \\
\hline
\end{tabular}
\label{tab:PET_usage}
\end{table}

\subsubsection{Use of Privacy-Enhancing Technologies (PETs)}
\label{sec:pets}

Using PETs is another critical component of privacy protection, that allows better protection and maintenance of data privacy against outside threats. 
We asked the SD team, who are the main users of PETs, if they used any PETs, and if so to list them. 
Out of the 198 participants, 2 did not respond. 111 of them (56.63\%) mentioned that they use some PETs while 36 (18.37\%) do not. About 25\% are unsure. These results are almost aligned with the degree of familiarity and usage of PbD. There was an increase ($\sim$10\%) in PETs familiarity and/or usage in comparison to PbD, which shows that these technologies are more common and tangible for developers, especially those related to encryption and access control.
We grouped responses into 6 categories shown in Table~\ref{tab:PET_usage} (definitions in Appendix \ref{sec:appendix_qualitative}). Encryption and access control, which are primarily security-focused, were the most common, followed by anonymization methods and differential privacy. 

Lastly, we investigated the correlations between the PETs' familiarity and the company size (\textbf{H4a}), confidence in security and privacy measures (\textbf{H4b}), and education level (\textbf{H4c}) (see Appendix~\ref{sec:petshypotheses}). 
With the adjusted p-value, we find no correlations 
($p-value = 0.254$, $0.529$, and $0.704$, respectively). 

\begin{mdframed}[backgroundcolor=darkgray!20,linewidth=0pt,innerleftmargin=10pt,innerrightmargin=10pt,skipabove=5pt]
PETs are slightly more commonly used than PbD strategies. However, there is still a gap in their familiarity, where more than 40\% of developers do not use them or are unsure of their usage. The most commonly used PETs are more security-oriented concepts, than privacy.
\end{mdframed}

\begin{table}[t]
\centering
\caption{Frequency of Usage of the Developers' Forums}
\begin{tabular}{|l|c|c|c|c|c|}
\hline
\textbf{Forums} & \textbf{Never} & \textbf{Rarely} & \textbf{1-3/M} & \textbf{1-3/W} & \textbf{Daily} \\ 
\hline
\textbf{SO} & 13.1\% & 17.1\% & 26.1\% & 24.1\% &  19.6\% \\ 
\hline
\textbf{GitHub} & 18.4\% & 23.9\% & 23.4\% & 19.9\% &  14.4\%\\ 
\hline
\textbf{Reddit} & 30.5\% & 35.0\% & 20.0\% & 10.5\% &  4.0\% \\ 
\hline
\textbf{Quora} & 54.5\% & 27.0\% & 12.5\% & 5.5\% &  0.5\%\\ 
\hline
\end{tabular}
\label{table:forum-table}
\end{table}

\subsubsection{Developers' Sources for Privacy Information}
\label{sec:privacyknowledge}

As discussed in Section \ref{sec:related-works}, developers sometimes seek privacy-related guidance on forums, such as Reddit or Stack Overflow (SO). We asked the SD teams how often they use various developers' forums 
for their privacy-related questions. 
Table \ref{table:forum-table} shows the distribution of the responses and their frequencies. $\sim$70\% and $\sim$58\% of the respondents use SO and GitHub \emph{at least 1-3 times per month}, while for Reddit and Quora, this number is about 34.5\% and 18.5\%. About 57\% of the respondents find these forums very or extremely useful, while less than 6\% find them not useful at all. In cases where they do not find the answer on these forums, the SD team discusses their questions with the security or privacy experts, asks their teammates, or uses AI tools. In Appendix~\ref{sec:appendix_devforums}, we provide a more detailed analysis regarding the usage of the forums.

\begin{mdframed}[backgroundcolor=darkgray!20,linewidth=0pt,innerleftmargin=10pt,innerrightmargin=10pt,skipabove=5pt]
Developers often seek privacy-related information from online forums, where more than 50\% of participants use either Stack Overflow or GitHub at least 1-3 times per month and they find these forums useful. 
\end{mdframed}

\label{sec:findings}

\section{Location Analysis} 
 
Our large-scale survey has responses from the US (189 responses) and non-US (173 responses from 22 countries: EU+UK, South Africa, Canada (CA), Mexico, and Chile), enabling us to examine differences in perceptions, experiences, and behaviors. 
We group the countries into three regions based on their similarities in privacy regulations: US+CA (192), EU+UK (132), and `Other' countries (38). 
To evaluate the difference in \textit{perception}, we examine whether participants' location correlates with their confidence in privacy and security measures  (\textbf{H6a} in Appendix~\ref{sec:appendix_locationanalysis}) and the presence of a CPO (\textbf{H6b}). Both hypotheses do not hold ($p-values$ are 0.0567 and 0.6470). Table \ref{table:cpo-location} shows the presence of a CPO across the three regions. The percentage of `Yes' is almost equal between US+CA, EU+UK, and the `Other' countries, while slightly more US+CA participants mentioned ``no CPO'' than elsewhere. 
This is not surprising since GDPR, the UK Data Protection Act of 2018, and the US HIPAA (Art.164.530) all require having a privacy officer or officials in a similar role. 


\begin{table}[b]
\centering
\caption{Distribution of Location-based CPO Presence}
\begin{tabular}{|l|c|c|c|c|}
\hline
\textbf{Locations} & \textbf{Yes} & \textbf{No} & \textbf{Unsure} & \textbf{Others} \\ 
\hline
\textbf{US+CA} & 43.7\% & 41.5\% & 14.1\% & 0.7\% \\ 
\hline
\textbf{EU+UK} & 41.7\% & 36.1\% & 20.3\% & 1.9\% \\ 
\hline
\textbf{Other Countries} & 43.5\% & 30.4\% & 26.1\% & 0\% \\ 
\hline
\end{tabular}
\label{table:cpo-location}
\end{table}

We evaluated whether there is a significant difference between participants' \textit{experience} in the three regions regarding the creation of PIA (\textbf{H6c}) and the number of privacy breaches (\textbf{H6d}). With $p-value$ 0.7724, we find no correlation for PIA. However, there is a correlation between the regions and privacy breaches ($p-value =$ 0.0010). We also analyzed the \textit{privacy behaviors} in the three regions concerning familiarity with PbD (\textbf{H6e}) and usage of PETs (\textbf{H6f}). With $p-values$ 0.3120 and 0.8588, we do not find any correlation that suggests that usage of PETs and PbD are equally (un)common in all regions. 
Since participants are from regions governed by different privacy laws, we investigated their familiarity with CCPA~\cite{CCPA} (\textbf{H6g}) and GDPR~\cite{GDPR} (\textbf{H6h}). As expected, we find a significant correlation between the participant's familiarity with the two regulations ($p-values$ are $<0.0001$ and $0.0009$ respectively). 
Due to the global reach of many apps, SDLC teams are responsible for complying with various regulations.
We further evaluated the responses to the familiarity with each regulation in various regions. We combined the responses given for at least \emph{somewhat familiarity} (i.e., somewhat, very, extremely familiar) and found that participants in the US+CA are most familiar with HIPAA while the rest are most familiar with GDPR. Those from `Other' countries are also more familiar with the US regulations than those residing in the EU+UK. Table \ref{table:regulations-familiairity} shows the distribution. 

\begin{table}[tb]
\centering
\caption{Distribution of Regulations Familiarity}
\begin{tabular}{|l|c|c|c|c|c|}
\hline
\small\textbf{Location} & \small\textbf{GDPR} & \small\textbf{HIPAA} & \small\textbf{COPPA} & \small\textbf{CCPA} & \small\textbf{CPRA} \\ 
\hline
\textbf{US+CA} & 71\% & 84\% & 53\% & 48\% &  44\% \\ 
\hline
\small\textbf{EU+UK} & 89\% & 37\% & 38\% & 11\% &  9\% \\ 
\hline
\small\textbf{Others} & 69\% & 51\% & 57\% & 29\% &  29\% \\ 
\hline
\end{tabular}
\label{table:regulations-familiairity}
\end{table}

\label{sec:location-analysis}

\section{Discussion}

\noindent\textbf{Summary of Findings} 
Concerning \emph{privacy perception}, our survey identifies that the majority of the participants define privacy in terms of control over personal information and disclose only when needed, or in terms of security. In other research \cite{Horstmann2024ThoseTA, tahaei2021privacy}, data protection and security were the most common definitions. Having a CPO or a similar role positively impacts confidence in protecting users' data. However, we found out that a sizable portion of the participants are unaware of such a role in their company, which may lead to ineffectiveness in utilizing privacy tools or reducing privacy breaches. Lack of proper communication among various roles is a challenge that other research also identified~\cite{Horstmann2024ThoseTA, tahaei2021privacy}. Our findings also align with 
\cite{balebako2014privacy} and 
\cite{Horstmann2024ThoseTA}, which observed a correlation between company size and having a CPO. However, we did not observe significant location-based differences in these perceptions. This is interesting but not surprising, since GDPR, HIPAA, the UK Data Protection Act, and Protection of Personal Information Act (POPIA) all require a CPO or similar roles. Several of our US participants mentioned (in Q27) that they collect Protected Health Information (PHI), 
which falls under HIPAA; e.g., one participant says \emph{``Health related data about people involved with our insurance companies''}. The extensive privacy requirements from these regulations likely explain why we observed no significant geographical differences in terms of participants' confidence, familiarity with PbD, and the usage of PETS. 

In terms of \emph{privacy experience}, most participants rely on legal experts to help create privacy policies; unlike \cite{balebako2014privacy} where creating a privacy policy was not the priority. Our study also shows that participants are primarily concerned about multi-jurisdictional compliance. Most of them are not involved in creating a PIA. The majority of those involved believe a PIA should be created during the planning or analysis phases; 
this is almost similar to findings in \cite{hadar2018privacy, Horstmann2024ThoseTA}. Our participants emphasized the importance of detailed documentation regarding the data lifecycle, as well as using encryption and access control tools to protect the confidentiality and integrity of data. Interestingly, the QA teams rely more than others on security, privacy, and legal experts to implement and enforce privacy and security rules. Other studies did not examine the privacy practices of QA roles, separately.

Regarding \emph{privacy behavior}, we identified that less than half of the participants are aware of PbD and an even smaller number use them. Similar to\cite{TahaeiLiVaniea+2022+114+131}, `hide', `minimize', `inform', and `control' are more commonly used. The usage of PETs is slightly more prevalent than PbD, but the focus is more on security practices, such as encryption and access control; similar to other research that found security concepts are more tangible \cite{hadar2018privacy, balebako2014privacy, tahaei2021privacy, Horstmann2024ThoseTA}. Anonymization techniques are not used frequently enough. 
We also find that although $\sim53\%$ of our participants are from the US+CA, most are more familiar with GDPR than US-based regulations such as COPPA and CCPA. ISec experts are among the most knowledgeable about various regulations, while QA teams are the least familiar. Other works focus mainly on GDPR and CCPA and do not explore details regarding participants'  familiarity \cite{hadar2018privacy, balebako2014privacy, tahaei2021privacy, Horstmann2024ThoseTA}. 
Most participants tend to seek answers to their privacy questions from developers' forums in addition to legal/policy experts; unlike \cite{balebako2014privacy} where they used `friends' or `social media'.  

\noindent\textbf{Research Directions}
Insights from the related work and our survey results highlight the need for approaches to operationalize PbD strategies and incorporate them into design and development. 
PbD patterns~\cite{privacypatterns-list, privacypatterns-github} provide detailed information about their usage and high-level solutions, but still lack implementation. Approaches that detect privacy behaviors in code \cite{Jain2022, jain2023} and further link them to patterns, or leverage automated code generation techniques to generate code from privacy patterns are yet to be explored. 

Our survey highlights software teams' challenges in creating accurate PIAs and privacy policies. Research directions that focus on automated approaches to detect the information types, privacy practices, and purposes pre- \cite{huang2023mobile} and post-development \cite{Jain2022, jain2023}, or to generate privacy statements from code \cite{jain2021prigen} could 
alleviate the challenges regarding accuracy, consistency, and compliance.  

Developers seek answers to their privacy-related questions from developers' forums, though increasingly  
use tools such as ChatGPT \cite{brown2020language, openai2023gpt}. However, these tools may not always provide accurate responses \cite{delile2023evaluating}. Developing methods to help translate developers' privacy-related questions into accurate privacy code snippets requires further attention \cite{feng-etal-2020-codebert}.


Our survey indicates that software teams face challenges in understanding and adhering to privacy regulations; thus, there is a need for approaches to help better understand such regulations, and establish and maintain compliance. 
However, most research focuses on detailed requirements analysis, not suitable for agile app development. Future studies could focus not only on automated extraction of legal/privacy requirements but also on generating (privacy-related) user stories to be used in agile development. Research directions on automated approaches to monitor compliance and nudge developers towards compliant approaches are also worth addressing~\cite{checks}. 


\noindent\textbf{Educational Takeaway}
Similar to other work \cite{balebako2014privacy, Horstmann2024ThoseTA, tahaei2021privacy}, our work shows the need for a more focused educational approach toward privacy in the SDLC. While currently, many courses emphasize security, it is important to tailor specific courses that include advanced privacy topics such as: regulations; the importance of PIA and other artifacts; challenges in privacy policy creation; and approaches such as PbD, differential privacy, and federated learning. This distinction between privacy from security is crucial since privacy encompasses a broad spectrum of concerns, including data handling, user consent, and transparency. 
Software teams should be equipped with educational modules and tools that foster and support life-long learning of dynamic privacy concepts. 
Nudging developers towards more privacy-preserving solutions through online support and tools is important. Balebako et al. \cite{balebako2014improving} suggest that with the right guidance, developers can be encouraged to prioritize privacy in their design and development processes. 

\label{sec:discussion}

\section{Conclusion} 
In this paper,  we examined 
privacy perceptions, practices, and behaviors of SDLC team members during software development. Our findings suggest a need for standardized privacy practices, educational awareness and implementation of PbD, and a privacy expert 
to promote privacy awareness and compliance. We identified gaps in privacy practices among software teams. Finally, we provide research and educational directions to reduce the challenges in implementing these practices.

In the future, we will extend our research to conduct a comparative analysis within the US states. We will also evaluate whether developers over-claim their expertise in a new study. We will look into how privacy is taught at educational institutes, both in computer science and at Law schools. 


\label{sec:conclusions}

\section*{Acknowledgments}
This research was funded by NSF Award \# 2238047.

\bibliographystyle{plain}
\bibliography{references}

\begin{thebibliography}{100}

\bibitem{alfadel2023empirical}
Mahmoud Alfadel, Diego~Elias Costa, and Emad Shihab.
\newblock Empirical analysis of security vulnerabilities in python packages.
\newblock {\em Empirical Software Engineering}, 28(3):59, 2023.

\bibitem{aljeraisy2021privacy}
Atheer Aljeraisy, Masoud Barati, Omer Rana, and Charith Perera.
\newblock Privacy laws and privacy by design schemes for the internet of
  things: A developer’s perspective.
\newblock {\em ACM Computing Surveys (CSUR)}, 54(5):1--38, 2021.

\bibitem{alomar2022developers}
Noura Alomar and Serge Egelman.
\newblock Developers say the darnedest things: Privacy compliance processes
  followed by developers of child-directed apps.
\newblock {\em Proc. on Privacy Enhancing Technologies}, 4(2022):24, 2022.

\bibitem{Amaral2021}
Orlando Amaral, Sallam Abualhaija, Mehrdad Sabetzadeh, and Lionel Briand.
\newblock A model-based conceptualization of requirements for compliance
  checking of data processing against gdpr.
\newblock In {\em 2021 IEEE 29th Int. Requirements Engineering Conf. Workshops
  (REW)}, pages 16--20, 2021.

\bibitem{arizon2021understanding}
Renana Arizon-Peretz, Irit Hadar, Gil Luria, and Sofia Sherman.
\newblock Understanding developers’ privacy and security mindsets via climate
  theory.
\newblock {\em Empirical Software Engineering}, 26:1--43, 2021.

\bibitem{balebako2014improving}
Rebecca Balebako and Lorrie Cranor.
\newblock Improving app privacy: Nudging app developers to protect user
  privacy.
\newblock {\em IEEE Security \& Privacy}, 12(4):55--58, 2014.

\bibitem{balebako2014privacy}
Rebecca Balebako, Abigail Marsh, Jialiu Lin, Jason~I Hong, and Lorrie~Faith
  Cranor.
\newblock The privacy and security behaviors of smartphone app
  developers.(2014).
\newblock {\em DOI: http://dx. doi. org/10.1184}, 1, 2014.

\bibitem{bamberger2011privacy}
Kenneth~A Bamberger and Deirdre~K Mulligan.
\newblock Privacy on the books and on the ground.
\newblock {\em Stanford Law Review}, pages 247--315, 2011.

\bibitem{Bednar2019}
Kathrin Bednar, Sarah Spiekermann, and Marc Langheinrich.
\newblock Engineering privacy by design: Are engineers ready to live up to the
  challenge?
\newblock {\em The Information Society}, 35(3):122--142, 2019.

\bibitem{bennett2017governance}
Colin~J Bennett and Charles~D Raab.
\newblock {\em The governance of privacy: Policy instruments in global
  perspective}.
\newblock Routledge, 2017.

\bibitem{bhatia2016privacy}
J.~Bhatia and T.D. et~al. Breaux.
\newblock Privacy risk in cybersecurity data sharing.
\newblock In {\em Proc. of the ACM on Workshop on ISCS}, pages 57--64, 2016.

\bibitem{breaux2014eddy}
Travis~D. Breaux, Hanan Hibshi, and Ashwini Rao.
\newblock Eddy, a formal language for specifying and analyzing data flow
  specifications for conflicting privacy requirements.
\newblock {\em Requirements Engineering}, 19(3):281--307, 2014.

\bibitem{breaux2015detecting}
Travis~D. Breaux, Daniel Smullen, and Hanan Hibshi.
\newblock Detecting repurposing and over-collection in multi-party privacy
  requirements specifications.
\newblock In {\em Requirements Engineering Conference (RE), 2015 IEEE 23rd
  International}, pages 166--175. IEEE, 2015.

\bibitem{breslow1970generalized}
Norman Breslow.
\newblock A generalized kruskal-wallis test for comparing k samples subject to
  unequal patterns of censorship.
\newblock {\em Biometrika}, 57(3):579--594, 1970.

\bibitem{brown2020language}
Tom Brown, Benjamin Mann, Nick Ryder, Melanie Subbiah, Jared~D Kaplan, Prafulla
  Dhariwal, Arvind Neelakantan, Pranav Shyam, Girish Sastry, Amanda Askell,
  et~al.
\newblock Language models are few-shot learners.
\newblock {\em Advances in neural information processing systems},
  33:1877--1901, 2020.

\bibitem{bu2020privacy}
Fei Bu, Nengmin Wang, Bin Jiang, and Huigang Liang.
\newblock “privacy by design” implementation: Information system
  engineers’ perspective.
\newblock {\em International Journal of Information Management}, 53:102124,
  2020.

\bibitem{cavoukian2009}
Ann Cavoukian.
\newblock Privacy by design - the 7 foundational principles implementation and
  mapping of fair information practices.
\newblock {\em {www.privacybydesign.ca}}, 2009.

\bibitem{checks}
{Checks}.
\newblock Simplify compliance with google.
\newblock \url{https://checks.google.com/}, 2024 (accessed Jun 6, 2024).

\bibitem{cobigo2020protecting}
Virginie Cobigo, Konrad Czechowski, Hajer Chalghoumi, Amelie Gauthier-Beaupre,
  Hala Assal, Jeffery Jutai, Karen Kobayashi, Amanda Grenier, and Fatoumata
  Bah.
\newblock Protecting the privacy of technology users who have cognitive
  disabilities: Identifying areas for improvement and targets for change.
\newblock {\em Journal of Rehabilitation and Assistive Technologies
  Engineering}, 7:2055668320950195, 2020.

\bibitem{colesky2016critical}
Michael Colesky, Jaap-Henk Hoepman, and Christiaan Hillen.
\newblock A critical analysis of privacy design strategies.
\newblock In {\em 2016 IEEE security and privacy workshops (SPW)}, pages
  33--40. IEEE, 2016.

\bibitem{dalela2021mixed}
Asmita Dalela, Saverio Giallorenzo, Oksana Kulyk, Jacopo Mauro, and Elda Paja.
\newblock A mixed-method study on security and privacy practices in danish
  companies.
\newblock {\em arXiv preprint arXiv:2104.04030}, 2021.

\bibitem{danezis2015privacy}
George Danezis, Josep Domingo-Ferrer, Marit Hansen, Jaap-Henk Hoepman,
  Daniel~Le Metayer, Rodica Tirtea, and Stefan Schiffner.
\newblock Privacy and data protection by design-from policy to engineering.
\newblock {\em arXiv preprint arXiv:1501.03726}, 2015.

\bibitem{danilova2021you}
Anastasia Danilova, Alena Naiakshina, Stefan Horstmann, and Matthew Smith.
\newblock Do you really code? designing and evaluating screening questions for
  online surveys with programmers.
\newblock In {\em 2021 IEEE/ ACM 43rd International Conference on Software
  Engineering (ICSE)}, pages 537--548. IEEE, 2021.

\bibitem{delile2023evaluating}
Zack Delile, Sean Radel, Joe Godinez, Garrett Engstrom, Theo Brucker, Kenzie
  Young, and Sepideh Ghanavati.
\newblock Evaluating privacy questions from stack overflow: Can chatgpt
  compete?
\newblock In {\em 2023 IEEE 31st International Requirements Engineering
  Conference Workshops (REW)}, pages 239--244. IEEE, 2023.

\bibitem{dwork2006differential}
Cynthia Dwork.
\newblock Differential privacy.
\newblock In {\em International colloquium on automata, languages, and
  programming}, pages 1--12. Springer, 2006.

\bibitem{ekambaranathan2021money}
Anirudh Ekambaranathan, Jun Zhao, and Max Van~Kleek.
\newblock “money makes the world go around”: Identifying barriers to better
  privacy in children’s apps from developers’ perspectives.
\newblock In {\em Proceedings of the 2021 CHI Conference on Human Factors in
  Computing Systems}, pages 1--15, 2021.

\bibitem{GDPR}
{European Union}.
\newblock The eu general data protection regulation (gdpr).
\newblock \url{http://www.eugdpr.org/}, 2024 (accessed February 10, 2024).

\bibitem{ezzini2021}
Saad Ezzini, Sallam Abualhaija, Chetan Arora, Mehrdad Sabetzadeh, and Lionel~C.
  Briand.
\newblock Using domain-specific corpora for improved handling of ambiguity in
  requirements.
\newblock In {\em 2021 IEEE/ACM 43rd International Conference on Software
  Engineering (ICSE)}, pages 1485--1497, 2021.

\bibitem{coppa}
{Federal Trade Commission}.
\newblock Children’s online privacy protection rule; final rule.
\newblock \url{http://tinyurl.com/5fh55th2}, 2024 (accessed Feb 12, 2024).

\bibitem{feng-etal-2020-codebert}
Zhangyin Feng, Daya Guo, Duyu Tang, Nan Duan, Xiaocheng Feng, Ming Gong, Linjun
  Shou, Bing Qin, Ting Liu, Daxin Jiang, and Ming Zhou.
\newblock {C}ode{BERT}: A pre-trained model for programming and natural
  languages.
\newblock In {\em Findings of the Association for Computational Linguistics:
  EMNLP 2020}, pages 1536--1547. ACL, 2020.

\bibitem{GAP07}
Sepideh Ghanavati, Daniel Amyot, and Liam Peyton.
\newblock Towards a {{Framework}} for {{Tracking Legal Compliance}} in
  {{Healthcare}}.
\newblock In John Krogstie, Andreas Opdahl, and Guttorm Sindre, editors, {\em
  Advanced {{Information Systems Engineering}}}, pages 218--232. {Springer},
  2007.

\bibitem{GAP09}
Sepideh Ghanavati, Daniel Amyot, and Liam Peyton.
\newblock Compliance analysis based on a goal-oriented requirement language
  evaluation methodology.
\newblock In {\em 2009 17th IEEE International Requirements Engineering
  Conference}, pages 133--142. IEEE, 2009.

\bibitem{GAR14}
Sepideh Ghanavati, Daniel Amyot, and Andr{\'e} Rifaut.
\newblock Legal {{Goal}}-oriented {{Requirement Language}} ({{Legal GRL}}) for
  {{Modeling Regulations}}.
\newblock In {\em Proc. of the 6th {{International Workshop}} on {{Modeling}}
  in {{Software Engineering}}}, pages 1--6, New York, NY, USA, 2014. {ACM}.

\bibitem{gorla2014checking}
Alessandra Gorla, Ilaria Tavecchia, Florian Gross, and Andreas Zeller.
\newblock Checking app behavior against app descriptions.
\newblock In {\em Proc. of the 36th Int. Conference on Software Engineering},
  pages 1025--1035, 2014.

\bibitem{CCPA}
{Government of California}.
\newblock California consumer privacy act (ccpa).
\newblock \url{https://oag.ca.gov/privacy/ccpa}, 2022 (accessed July 20, 2022).

\bibitem{Green2016DevelopersAN}
Matthew Green and Matthew Smith.
\newblock Developers are not the enemy!: The need for usable security apis.
\newblock {\em IEEE Security \& Privacy}, 14:40--46, 2016.

\bibitem{greenwood1996guide}
Priscilla~E Greenwood and Michael~S Nikulin.
\newblock {\em A guide to chi-squared testing}, volume 280.
\newblock John Wiley \& Sons, 1996.

\bibitem{gustavsson2020assessment}
Sara Gustavsson.
\newblock An assessment of privacy by design as a stipulation in gdpr.
\newblock 2020.

\bibitem{Hadar2018}
Irit Hadar, Tomer Hasson, Oshrat Ayalon, Eran Toch, Michael Birnhack, Sofia
  Sherman, and Arod Balissa.
\newblock Privacy by designers: Software developers' privacy mindset.
\newblock {\em Journal of Empirical Software Engineering}, 23(1):259–289,
  February 2018.

\bibitem{hadar2018privacy}
Irit Hadar, Tomer Hasson, Oshrat Ayalon, Eran Toch, Michael Birnhack, Sofia
  Sherman, and Arod Balissa.
\newblock Privacy by designers: software developers’ privacy mindset.
\newblock {\em Empirical Software Engineering}, 23(1):259--289, 2018.

\bibitem{Harkous2022}
Hamza Harkous, Sai~Teja Peddinti, Rishabh Khandelwal, Animesh Srivastava, and
  Nina Taft.
\newblock Hark: A deep learning system for navigating privacy feedback at
  scale.
\newblock In {\em IEEE Symp. on Security and Privacy}, 2022.

\bibitem{HE2019259}
Yongzhong He, Xuejun Yang, Binghui Hu, and Wei Wang.
\newblock Dynamic privacy leakage analysis of android third-party libraries.
\newblock {\em Journal of Information Security and Applications}, 46:259--270,
  2019.

\bibitem{hipaa}
US~Department Health and Human Services.
\newblock {The Health Insurance Portability and Accountability Act (HIPAA)}.
\newblock \url{https://www.hhs.gov/hipaa/index.html}, 2024 (accessed Feb 10,
  2024).

\bibitem{Hoepman2014PrivacyDS}
J.~Hoepman.
\newblock Privacy design strategies (extended abstract).
\newblock 2014.

\bibitem{hoepman2018making}
J-H Hoepman.
\newblock Making privacy by design concrete.
\newblock 2018.

\bibitem{hoofnagle2010different}
Chris~Jay Hoofnagle, Jennifer King, Su~Li, and Joseph Turow.
\newblock How different are young adults from older adults when it comes to
  information privacy attitudes and policies?
\newblock {\em Available at SSRN 1589864}, 2010.

\bibitem{hopemanprivacy}
Jaap-Henk Hopeman and Marc~van Lieshout.
\newblock Privacy: a fundamental right.

\bibitem{Horstmann2024ThoseTA}
Stefan~Albert Horstmann, Samuel Domiks, Marco Gutfleisch, Mindy Tran, Yasemin
  Acar, Veelasha Moonsamy, and Alena Naiakshina.
\newblock "those things are written by lawyers, and programmers are reading
  that." mapping the communication gap between software developers and privacy
  experts.
\newblock {\em Proc. Priv. Enhancing Technol.}, 2024:151--170, 2024.

\bibitem{huang2023mobile}
Tianjian Huang, Vaishnavi Kaulagi, Mitra~Bokaei Hosseini, and Travis Breaux.
\newblock Mobile application privacy risk assessments from user-authored
  scenarios.
\newblock In {\em Proceedings of the 31st IEEE International Requirements
  Engineering Conference}, pages 1--12. IEEE, 2023.

\bibitem{IAPP-Taxonomy}
{International Association of Privacy Professionals}.
\newblock Taxonomy of privacy.
\newblock \url{https://iapp.org/resources/article/a-taxonomy-of-privacy/a},
  2024 (accessed June 1, 2024).

\bibitem{iwaya2023privacy}
Leonardo~Horn Iwaya, Muhammad~Ali Babar, and Awais Rashid.
\newblock Privacy engineering in the wild: Understanding the practitioners'
  mindset, organisational aspects, and current practices.
\newblock {\em IEEE Transactions on Software Engineering}, 2023.

\bibitem{jain2023atlas}
Akshath Jain, David Rodriguez, Jose~M del Alamo, and Norman Sadeh.
\newblock Atlas: Automatically detecting discrepancies between privacy policies
  and privacy labels.
\newblock {\em arXiv preprint arXiv:2306.09247}, 2023.

\bibitem{jain2023}
Vijayanta Jain, Sepideh Ghanavati, Sai~Teja Peddinti, and Collin McMillan.
\newblock Towards fine-grained localization of privacy behaviors.
\newblock In {\em IEEE 8th European Symposium on Security and Privacy}, pages
  258--277, 2023.

\bibitem{jain2021prigen}
Vijayanta Jain, Sanonda~Datta Gupta, Sepideh Ghanavati, and Sai~Teja Peddinti.
\newblock Prigen: Towards automated translation of android applications’ code
  to privacy captions.
\newblock In {\em Int. Conference on Research Challenges in Information
  Science}, pages 142--151. Springer, 2021.

\bibitem{Jain2022}
Vijayanta Jain, Sanonda~Datta Gupta, Sepideh Ghanavati, Sai~Teja Peddinti, and
  Collin McMillan.
\newblock Pact: Detecting and classifying privacy behavior of android
  applications.
\newblock In {\em Proc. of the 15th ACM Conf. on Security and Privacy in
  Wireless and Mobile Networks}, WiSec '22, page 104–118. ACM, 2022.

\bibitem{kaur2022recruit}
Harjot Kaur, Sabrina Amft, Daniel Votipka, Yasemin Acar, and Sascha Fahl.
\newblock Where to recruit for security development studies: Comparing six
  software developer samples.
\newblock In {\em 31st USENIX Security Symposium (USENIX Security 22)}, pages
  4041--4058, 2022.

\bibitem{khandelwal2023unpacking}
Rishabh Khandelwal, Asmit Nayak, Paul Chung, and Kassem Fawaz.
\newblock Unpacking privacy labels: A measurement and developer perspective on
  google's data safety section.
\newblock {\em arXiv preprint arXiv:2306.08111}, 2023.

\bibitem{kruskal1952use}
William~H Kruskal and W~Allen Wallis.
\newblock Use of ranks in one-criterion variance analysis.
\newblock {\em Journal of the American statistical Association},
  47(260):583--621, 1952.

\bibitem{li2021developers}
Tianshi Li, Elizabeth Louie, Laura Dabbish, and Jason~I Hong.
\newblock How developers talk about personal data and what it means for user
  privacy: A case study of a developer forum on reddit.
\newblock {\em Proceedings of the ACM on Human-Computer Interaction},
  4(CSCW3):1--28, 2021.

\bibitem{liu2018large}
Xueqing Liu, Yue Leng, Wei Yang, Wenyu Wang, Chengxiang Zhai, and Tao Xie.
\newblock A large-scale empirical study on android runtime-permission rationale
  messages.
\newblock In {\em 2018 IEEE Symposium on Visual Languages and Human-Centric
  Computing (VL/HCC)}, pages 137--146. IEEE, 2018.

\bibitem{macleod2017documenting}
Laura MacLeod, Andreas Bergen, and Margaret-Anne Storey.
\newblock Documenting and sharing software knowledge using screencasts.
\newblock {\em Empirical Software Engineering}, 22:1478--1507, 2017.

\bibitem{PETsMatrix-ENISA}
European Union Agency~For Network and Information Security.
\newblock Pets controls matrix a systematic approach for assessing online and
  mobile privacy tools.
\newblock 2016.

\bibitem{Alomar2022}
Serge~Egelman Noura Alomar~and and Jordan~L. Fischer.
\newblock Developers say the darnedest things: Privacy compliance processes
  followed by developers of child-directed apps.
\newblock {\em Proceedings on Privacy Enhancing Technologies}, 2022(4), 2022.

\bibitem{ferpa}
US~Department of~Education.
\newblock {The Family Educational Rights and Privacy Act (FERPA)}.
\newblock \url{https://www2.ed.gov/policy/gen/guid/fpco/ferpa/index.html}, 2024
  (accessed Feb 10, 2024).

\bibitem{okoyomon2019ridiculousness}
Ehimare Okoyomon, Nikita Samarin, Primal Wijesekera, Amit Elazari Bar~On,
  Narseo Vallina-Rodriguez, Irwin Reyes, {\'A}lvaro Feal, and Serge Egelman.
\newblock On the ridiculousness of notice and consent: Contradictions in app
  privacy policies.
\newblock 2019.

\bibitem{unctad_2021}
United Nations~Conference on~Trade and Development.
\newblock Data protection and privacy legislation worldwide.
\newblock \url{https://tinyurl.com/puev83dt}, 2021.

\bibitem{openai2023gpt}
R~OpenAI.
\newblock Gpt-4 technical report.
\newblock {\em arXiv}, pages 2303--08774, 2023.

\bibitem{palan2018prolific}
Stefan Palan and Christian Schitter.
\newblock Prolific. ac—a subject pool for online experiments.
\newblock {\em Journal of Behavioral and Experimental Finance}, 17:22--27,
  2018.

\bibitem{pandita2013whyper}
Rahul Pandita, Xusheng Xiao, Wei Yang, William Enck, and Tao Xie.
\newblock $\{$WHYPER$\}$: Towards automating risk assessment of mobile
  applications.
\newblock In {\em Presented as part of the 22nd $\{$USENIX$\}$ Security
  Symposium ($\{$USENIX$\}$ Security 13)}, pages 527--542, 2013.

\bibitem{parsons2023understanding}
Jonathan Parsons, Michael Schrider, Oyebanjo Ogunlela, and Sepideh Ghanavati.
\newblock Understanding developers privacy concerns through reddit thread
  analysis.
\newblock {\em Joint Proc. of REFSQ-2023 Workshops, Doctoral Symposium, Posters
  \& Tools Track and Journal Early Feedback co-located with the 28th Int. Conf.
  on Requirements Engineering: Foundation for Software Quality (REFSQ 2023),
  Barcelona, Catalunya}, 2023.

\bibitem{qu2014autocog}
Zhengyang Qu, Vaibhav Rastogi, Xinyi Zhang, Yan Chen, Tiantian Zhu, and Zhong
  Chen.
\newblock Autocog: Measuring the description-to-permission fidelity in android
  applications.
\newblock In {\em Proceedings of the 2014 ACM SIGSAC Conference on Computer and
  Communications Security}, pages 1354--1365, 2014.

\bibitem{rodriguez2023comparing}
David Rodriguez, Akshath Jain, Jose~M Del~Alamo, and Norman Sadeh.
\newblock Comparing privacy label disclosures of apps published in both the app
  store and google play stores.
\newblock In {\em IEEE European Symp. on Security and Privacy Workshops}, pages
  150--157, 2023.

\bibitem{SDLC2010}
Nayan~B. Ruparelia.
\newblock Software development lifecycle models.
\newblock {\em SIGSOFT Softw. Eng. Notes}, 35(3):8–13, 2010.

\bibitem{serafini2023recruitment}
Raphael Serafini, Marco Gutfleisch, Stefan~Albert Horstmann, and Alena
  Naiakshina.
\newblock On the recruitment of company developers for security studies:
  results from a qualitative interview study.
\newblock In {\em 19th Symposium on Usable Privacy and Security}, pages
  321--340, 2023.

\bibitem{Bonferroni}
J~P Shaffer.
\newblock Multiple hypothesis testing.
\newblock {\em Annual Review of Psychology}, 46(1):561--584, 1995.

\bibitem{shen2011privacy}
Yun Shen and Siani Pearson.
\newblock Privacy enhancing technologies: A review.
\newblock {\em Hewlet Packard Development Company. Disponible en https://bit.
  ly/3cfpAKz}, 2011.

\bibitem{slavin2016pvdetector}
Rocky Slavin, Xiaoyin Wang, Mitra~Bokaei Hosseini, James Hester, Ram Krishnan,
  Jaspreet Bhatia, Travis~D Breaux, and Jianwei Niu.
\newblock Pvdetector: a detector of privacy-policy violations for android apps.
\newblock In {\em 2016 IEEE/ACM Int. Conf. on Mobile Software Engineering and
  Systems (MOBILESoft)}, pages 299--300, 2016.

\bibitem{slavin2016toward}
Rocky Slavin, Xiaoyin Wang, Mitra~Bokaei Hosseini, James Hester, Ram Krishnan,
  Jaspreet Bhatia, Travis~D Breaux, and Jianwei Niu.
\newblock Toward a framework for detecting privacy policy violations in android
  application code.
\newblock In {\em Proceedings of the 38th International Conference on Software
  Engineering}, pages 25--36, 2016.

\bibitem{solove2006taxonomy}
Daniel~J Solove.
\newblock A taxonomy of privacy.
\newblock {\em University of Pennsylvania law review}, pages 477--564, 2006.

\bibitem{cranor2009}
Sarah Spiekermann and Lorrie~Faith Cranor.
\newblock Engineering privacy.
\newblock {\em IEEE Transactions on Software Engineering}, 35(1):67--82, 2009.

\bibitem{spiekermann2018inside}
Sarah Spiekermann, Jana Korunovska, and Marc Langheinrich.
\newblock Inside the organization: Why privacy and security engineering is a
  challenge for engineers.
\newblock {\em Proceedings of the IEEE}, 107(3):600--615, 2018.

\bibitem{spiekermann2018understanding}
Sarah Spiekermann-Hoff, Jana Korunovska, and Marc Langheinrich.
\newblock Understanding engineers' drivers and impediments for ethical system
  development: The case of privacy and security engineering.
\newblock 2018.

\bibitem{sweeney2002k}
Latanya Sweeney.
\newblock k-anonymity: A model for protecting privacy.
\newblock {\em Int. journal of uncertainty, fuzziness and knowledge-based
  systems}, 10(05):557--570, 2002.

\bibitem{tahaei2022privacy}
Mohammad Tahaei, Julia Bernd, and Awais Rashid.
\newblock Privacy, permissions, and the health app ecosystem: A stack overflow
  exploration.
\newblock In {\em Proc. of the 2022 European Symposium on Usable Security},
  pages 117--130, 2022.

\bibitem{tahaei2021privacy}
Mohammad Tahaei, Alisa Frik, and Kami Vaniea.
\newblock Privacy champions in software teams: Understanding their motivations,
  strategies, and challenges.
\newblock In {\em Proceedings of the 2021 CHI Conference on Human Factors in
  Computing Systems}, pages 1--15, 2021.

\bibitem{Tahaei2021Springer}
Mohammad Tahaei, Adam Jenkins, Kami Vaniea, and Maria Wolters.
\newblock ``i don't know too much about it'': On the security mindsets of
  computer science students.
\newblock In Thomas Gro{\ss} and Theo Tryfonas, editors, {\em Socio-Technical
  Aspects in Security and Trust}, pages 27--46, Cham, 2021. Springer
  International Publishing.

\bibitem{TahaeiLiVaniea+2022+114+131}
Mohammad Tahaei, Tianshi Li, and Kami Vaniea.
\newblock Understanding privacy-related advice on stack overflow.
\newblock {\em Proceedings on Privacy Enhancing Technologies},
  2022(2):114--131, 2022.

\bibitem{tahaei2019survey}
Mohammad Tahaei and Kami Vaniea.
\newblock A survey on developer-centred security.
\newblock In {\em 2019 IEEE European Symposium on Security and Privacy
  Workshops (EuroS\&PW)}, pages 129--138. IEEE, 2019.

\bibitem{tahaei2021developers}
Mohammad Tahaei and Kami Vaniea.
\newblock “developers are responsible”: What ad networks tell developers
  about privacy.
\newblock In {\em Extended Abstracts in CHI Conf. on Human Factors in Computing
  Systems}, pages 1--11, 2021.

\bibitem{Tahaei-CHI2022}
Mohammad Tahaei and Kami Vaniea.
\newblock Recruiting participants with programming skills: A comparison of four
  crowdsourcing platforms and a cs student mailing list.
\newblock In {\em CHI Conference on Human Factors in Computing Systems}, CHI
  '22. ACM, 2022.

\bibitem{tahaei2020understanding}
Mohammad Tahaei, Kami Vaniea, and Naomi Saphra.
\newblock Understanding privacy-related questions on stack overflow.
\newblock In {\em Proceedings of the 2020 CHI conference on human factors in
  computing systems}, pages 1--14, 2020.

\bibitem{FTC}
{The Federal Trade Commission}.
\newblock Privacy and security enforcement.
\newblock 2024 (accessed Feb 10, 2024).

\bibitem{privacypatterns-github}
{UC - Berkeley - School of Information}.
\newblock Privacy patterns - collaborative development of privacy software
  design patterns.
\newblock \url{https://github.com/privacypatterns}, 2024 (accessed Feb. 10,
  2024).

\bibitem{privacypatterns-list}
{UC Berkeley - School of Information}.
\newblock Privacy patterns org.
\newblock \url{https://privacypatterns.org/}, 2024 (accessed February 10,
  2024).

\bibitem{globalbreaches}
{Varonis}.
\newblock 84 must-know data breach statistics for 2023.
\newblock \url{https://www.varonis.com/blog/data-breach-statistics}, (accessed
  Feb. 10, 2024).

\bibitem{yu2016can}
L.~Yu and X.~et~al. Lou.
\newblock Can we trust the privacy policies of android apps?
\newblock In {\em 46th Annual IEEE/IFIP Int. Conf. on (DSN)}, pages 538--549.
  IEEE, 2016.

\bibitem{Zhan2021}
Xian Zhan, Lingling Fan, Sen Chen, Feng We, Tianming Liu, Xiapu Luo, and Yang
  Liu.
\newblock Atvhunter: Reliable version detection of third-party libraries for
  vulnerability identification in android applications.
\newblock In {\em 2021 IEEE/ACM 43rd International Conference on Software
  Engineering (ICSE)}, pages 1695--1707, 2021.

\bibitem{zimmeck2021privacyflash}
Sebastian Zimmeck, Rafael Goldstein, and David Baraka.
\newblock Privacyflash pro: Automating privacy policy generation for mobile
  apps.
\newblock 2021.

\bibitem{zimmeck2019maps}
Sebastian Zimmeck, Peter Story, Daniel Smullen, Abhilasha Ravichander, Ziqi
  Wang, Joel~R Reidenberg, N~Cameron Russell, and Norman Sadeh.
\newblock Maps: Scaling privacy compliance analysis to a million apps.
\newblock {\em Proc. Priv. Enhancing Tech.}, 2019:66, 2019.

\bibitem{zimmer2008gaze}
Michael Zimmer.
\newblock The gaze of the perfect search engine: Google as an infrastructure of
  dataveillance.
\newblock In {\em Web search: Multidisciplinary perspectives}, pages 77--99.
  Springer, 2008.

\end{thebibliography}

\appendix

\section{Survey Questions}
\label{sec:appendix_survey}

Survey questions can be found here: \url{http://tinyurl.com/2p9n49e4}

\section{Participants' Demographic Information}
\label{sec:appendix_demographic}

Table \ref{table:demographics} below shows the various demographics of our participants.

\begin{table*}[t]
\centering
\footnotesize
\caption{Demographic Information about the Participants}
\begin{tabular}{|l|c|c|c|c|c|}
\hline
\textbf{Gender}  & Female (25.48\%) & Male (73.41\%) & Non-Binary (0.55\%) & Other (0.55\%) & PnS (0\%) 
  \\ 
\hline
\textbf{Age} & 18-25 (19.89\%) & 26-35 (45.86\%) & 36-45 (20.99\%) & 46-55 (8.84\%) & $>$55 (3.87\%) \\ 
\hline
\textbf{Education} & High school (10.22\%) & BSc. (61.05\%) & MSc. (22.10\%) &  PhD (1.66\%) & Other (3.87\%)  \\ 
\hline
\textbf{Degree}  & CS/ECE/DS (34.8\%) & IT (26.24\%) &  Business (11.05\%) & Other (24.04\%) & PnS (3.87\%) \\
\hline
\textbf{Company Size}  & 100+ emp. (50.00\%) & 50-100 (13.54\%) & 21-50  (12.43\%) & 11-20  (7.46\%) & 0-10 (16.57\%) \\
\hline
\end{tabular}
\label{table:demographics}
\end{table*}

\section{Details of Solove's Taxonomy}
\label{sec:appendix_solove}

Solove's Taxonomy and the mapping of subcategories.

\begin{table}[H]
\centering
\footnotesize
\caption{Solove's Categories and Subcategories}
\begin{tabular}{|p{3.5cm}|p{4cm}|}
\hline
\textbf{Main Category} & \textbf{Solove's Subcategories} \\
\hline
Information Collection & Surveillance, Interrogation \\
\hline
Information Processing & Aggregation, Identification, Insecurity, Secondary Use, Exclusion \\
\hline
Information Dissemination & Breach of Confidentiality, Disclosure, Exposure, Increased Accessibility, Blackmail, Appropriation, Distortion \\
\hline
Invasion & Intrusion, Decisional Interference \\
\hline
\end{tabular}
\label{table:soloves_categories}
\end{table}





\section{Confidence in Security \& Privacy Measures}
\label{sec:appendix-confidece}

The hypotheses list for the correlation between confidence in security and privacy measures and various factors are:

\noindent\textbf{-- H1a:} The size of the company correlates with confidence in privacy and security measures.

\noindent\textbf{-- H1b:} The participants' role at the company correlates to confidence in privacy and security measures.

\noindent\textbf{-- H1c:} The education level correlates to confidence in privacy and security measures.
  
\noindent\textbf{-- H1d:} The presence of a CPO or similar position correlates to confidence in privacy and security measures.

The p-value results of the Chi-Square tests are as follows:

\begin{table}[H]
\centering
\footnotesize
\caption{P-Value for Hypothesis \textbf{H1a} to \textbf{H1d}}
\begin{tabular}{|l|c|c|c|c|}
\hline
\textbf{} & \textbf{H1a} & \textbf{H1b} & \textbf{H1c} & \textbf{H1d}\\
\hline
\textbf{P-Value} & 0.494 & 0.654 & 0.570 & 0.0007\\
\hline
\end{tabular}
\label{table:hypothesis1}
\end{table}

\section{Presence of a CPO or a Similar Role}
\label{sec:appendix-cpo}

The participant's knowledge about the presence of a CPO in their company is as follows:

\begin{table}[H]
\centering
\footnotesize
\caption{Distribution of Knowledge about a CPO}
\begin{tabular}{|c|c|c|c|}
\hline
\textbf{Yes} & \textbf{No} & \textbf{Unsure} & \textbf{Others} \\ 
\hline
42.6\% & 38.4\% & 17.9\% & 1.1\% \\ 
\hline
\end{tabular}
\label{table:cpo}
\end{table}

The hypotheses list for the correlation between the presence of a CPO/a similar role and the PIA creation, familiarity with PETs, number of privacy breaches, and the company size are: 

  \noindent\textbf{-- H2a:} The creation of a PIA correlates to the presence of a CPO or similar position at the company.
  
   \noindent\textbf{-- H2b:} Familiarity with PETs correlates to the presence of a CPO or similar position at the company.
   
   
    \noindent\textbf{-- H2c:} The higher number of privacy breaches correlates to the presence of a CPO or similar position at the company. 
    
    \noindent\textbf{-- H2d:} The size of a company correlates to the presence of a CPO or similar position at the company.

The p-value results of the Chi-Square tests are as follows:

\begin{table}[H]
\centering
\footnotesize
\caption{P-Value for Hypothesis \textbf{H2a} to \textbf{H2d}}
\begin{tabular}{|l|c|c|c|c|}
\hline
\textbf{} & \textbf{H2a} & \textbf{H2b} & \textbf{H2c} & \textbf{H2d} \\
\hline
\textbf{P-Value} & 0.1005 & 0.008 & 0.359 & $<0.00001$ \\
\hline
\end{tabular}
\label{table:hypothesis2}
\end{table}

The distribution of how participants address their compliance questions:

\begin{table}[H]
\centering
\footnotesize
\caption{Distribution of Sources for Compliance Questions}
\begin{tabular}{|c|c|c|c|c|}
\hline
\textbf{Lawyer} & \textbf{CPO} & \textbf{Best Practices} & \textbf{Forums} & \textbf{Others} \\ 
\hline
24.2\% & 25.9\% & 23.1\% & 18.5\% & 8.3\% \\ 
\hline
\end{tabular}
\label{table:compliance-questions}
\end{table}






\section{The Creation of a PIA}
\label{sec:appendix-PIA}

The hypotheses list for the correlation between the creation of a PIA and the company size and confidence in privacy and security measures are:

  \noindent\textbf{-- H3a:} The size of the company correlates to the PIA creation.
  
  \noindent\textbf{-- H3b:} The participants' confidence in an organization's privacy and security measures correlates to the PIA creation.

The p-value results of the Chi-Square tests are as follows:

\begin{table}[H]
\centering
\footnotesize
\caption{P-Value for Hypothesis \textbf{H3a} to \textbf{H3b}}
\begin{tabular}{|l|c|c|}
\hline
\textbf{} & \textbf{H2a} & \textbf{H2b} \\
\hline
\textbf{P-Value} & $<0.00001$ & $<0.00001$ \\
\hline
\end{tabular}
\label{table:hypothesis3}
\end{table}

\begin{figure}[thbp]
    \centering
    \includegraphics[width=0.28\textwidth]{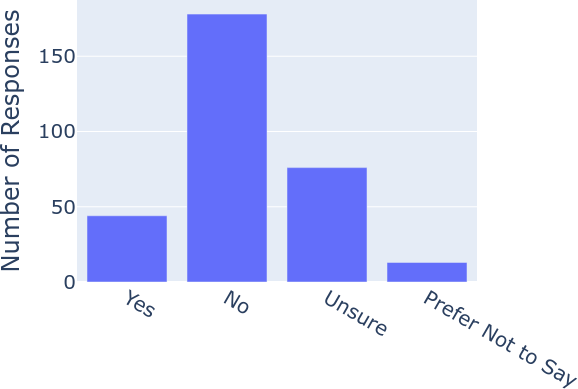}
    \caption{Distribution of Responses to the Creation of a PIA.
    }
    \label{fig:creation_PIA}
\end{figure}

The distribution of responses to the creation of a PIA in their company is shown in Figure~\ref{fig:creation_PIA}.

\section{Privacy by Design Approaches}
\label{sec:PbD-Appendix}

The distribution of participants who are familiar with PbD:

\begin{table}[H]
\centering
\footnotesize
\caption{Distribution of Familiarity with PbD Strategies}
\begin{tabular}{|l|c|c|c|c|}
\hline
\textbf{Role} & \textbf{Yes} & \textbf{No} & \textbf{Unsure} & \textbf{PnS} \\ 
\hline
\textbf{SD} & 91 (46\%) & 54 (27.3\%) & 49 (24.7\%) & 4 (2\%) \\ 
\hline
\end{tabular}
\label{table:PbDFamiliairty}
\end{table}

\section{Detailed Analysis of PETs' Familiarity}
\label{sec:petshypotheses}
The list of the hypotheses for the correlation between the usage of PETs and the size of the company, participants' confidence, and the presence of the CPO is as follows:

  \noindent\textbf{-- H4a:} The size of the company correlates to the use of PETs.
  
  \noindent\textbf{-- H4b:} The participant's confidence in an organization's privacy and security measures correlates to the use of PETs.
  
  \noindent\textbf{-- H4c:} The participant's education level correlates to the use of PETs.

Table \ref{table:hypothesis5} shows the results of the hypotheses analysis.

\begin{table}[H]
\centering
\footnotesize
\caption{P-Value for Hypothesis \textbf{H4a} to \textbf{H4c}}
\begin{tabular}{|c|c|c|c|}
\hline
\textbf{} & \textbf{H4a} & \textbf{H4b} & \textbf{H4c} \\
\hline
\textbf{P-Value} & 0.254 & 0.704 & 0.529\\
\hline
\end{tabular}
\label{table:hypothesis5}
\end{table}




\begin{table}[H]
\centering
\footnotesize
\caption{P-Value and H Value for Hypothesis \textbf{H5a} to \textbf{H5c}}
\begin{tabular}{|p{1.2cm}|p{1.2cm}|p{1.2cm}|p{1.2cm}|}
\hline
\textbf{} & \textbf{H5a} & \textbf{H5b} & \textbf{H5c} \\
\hline
\textbf{P-Value} & 0.04 & 0.17 & 0.08 \\
\hline
\textbf{H Value} & 4.03 & 7.80 & 9.83 \\
\hline
\end{tabular}
\label{table:hypothesis7}
\end{table}

\begin{table*}[htb]
\centering
\footnotesize
\caption{Categories of PETs}
\begin{tabular}{|p{3cm}|p{6.3cm}|p{6.5cm}|} 
\hline
\textbf{Categories of PETs} & \textbf{Definition in Literature \cite{PETsMatrix-ENISA, shen2011privacy}}
 & \textbf{Example from Our Survey} \\
\hline
\textbf{Encryption}  & A system of communication where the only people who can read the messages are the people communicating. 
 & We use encryption and a number of security features offered by the platform we implement. It is primarily the responsibility of the back-end programmers.
 \\ 
\hline
\textbf{Access Control/ Identity protection
}  & Deals with identifying individuals and controlling access to resources in a system.
 & We implement role-based access for the various features of our product as well as internally
 \\ 
\hline
\textbf{Anonymity and Pseudonymity
}  & Involves removing personally identifiable information (PII) to prevent individual users from being identified. Pseudonymity involves replacing identifiers with pseudonyms \cite{sweeney2002k}.
 & Data anonymization, our managers would be the primary users for that subject
 \\ 
\hline
\textbf{Differential Privacy}  & Involves adding noise to the data to protect individual user information while still providing useful insights. It is particularly useful in data analysis and machine learning applications. \cite{dwork2006differential}
 & We use encryption and al little bit of \textbf{differential privacy} where it is applicable and it varies from project to project with who is tasked with implementing these features.

 \\ 
\hline
\textbf{Secure Communication/ VPN
}  & Involves encrypting all communications within the software using standard protocols like HTTPS and SSL/TLS. 
 & All of our internal communication is done over an internal VPN, and all web access is done with https.
 \\ 
\hline
\textbf{Privacy-Enhance Anti Web Tracking}  & Involves blocking attempts of different types of trackers to monitor users’ online activity and personal data. 
 & -
 \\ 
\hline
\end{tabular}
\label{table:PETS_Taxonomy}
\end{table*}

\section{Factors Influencing Usage of Forums}
\label{sec:appendix_devforums}

To further evaluate the impact of the size of the company, familiarity with PETs, and the presence of a CPO on the usage of developer forums, we employed the Kruskal-Wallis test which is a non-parametric test that is used to compare two or more independent samples for statistically significant differences between groups \cite{kruskal1952use}. 
Below is the list of hypotheses for the frequency of the usage of the developers' forums:

  \noindent\textbf{-- H5a:} The size of the company correlates to the use of developer forums to ask privacy-related questions.
  
  \noindent\textbf{-- H5b:} The presence of a Chief Privacy Officer or similar position at a participant's organization correlates to the use of developer forums to ask privacy-related questions.
  
  \noindent\textbf{-- H5c:} Familiarity with PETs correlates to the use of developer forums to ask privacy-related questions.

As shown in Table \ref{table:hypothesis7} when comparing forum usage with the size of the company, a statistically significant difference was found between the groups ($H-Value = 4.03, p-value = 0.04$). However, no significant difference was noted when comparing forum usage with the presence of a Chief Privacy Officer (CPO) ($H-value = 9.83, p-value = 0.08$) or with the usage of Privacy Enhancing Technologies (PETs) ($H-value = 3.92, p-value = 0.56$). These findings suggest that only the size of the company is more likely to influence the frequency with which developers consult forums for privacy-related inquiries.




\section{Details for the Location Analysis}
\label{sec:appendix_locationanalysis}
Below is the list of hypotheses for location analysis.  

  \noindent\textbf{-- H6a:} The participants' confidence in their organization's privacy and security measures correlates to their region of origin.
 
  \noindent\textbf{-- H6b:} The presence of a CPO or similar position at a participant's organization correlates to their region of origin.
 
  \noindent\textbf{-- H6c:} The participants' creation of a PIA correlates to their region of origin.
  
  \noindent\textbf{-- H6d:} The participants' organization being a victim of a breach of privacy correlates to their region of origin.
  
  \noindent\textbf{-- H6e:} The participants' familiarity with PbD strategies correlates to their region of origin.
 
  \noindent\textbf{-- H6f:} The participants' use of PETs correlates to their region of origin.
  
  \noindent\textbf{-- H6g:} The participants' familiarity with the CCPA correlates to their region of origin.
 
  \noindent\textbf{-- H6h:} The participants' familiarity with the GDPR correlates to their region of origin.

\section{Qualitative Analysis Guidelines}
\label{sec:appendix_qualitative}

Table~\ref{table:PETS_Taxonomy} shows the different categories of PETs and Table~\ref{table:privacy_taxonomy} describes the privacy taxonomy, both of which were considered as guidelines for our qualitative analysis.

\begin{table*}[htbp]
\centering
\footnotesize
\caption{Taxonomy of Privacy}
\begin{tabular}{|p{1.8cm}|p{4.3cm}|p{4.5cm}|p{4.4cm}|} 
\hline
\textbf{Taxonomy of Privacy} & \textbf{Solove’s Definition~\cite{solove2006taxonomy}} 
& \textbf{Example from IAPP~\cite{IAPP-Taxonomy}} & \textbf{Example from Our Survey} \\
\hline
\textbf{Surveillance}  & Watching, listening to, or recording of an individual's activities
 & A website monitoring the cursor movements of a visitor while visiting the website.
 & Privacy is the ability to keep information or activities out of public knowledge \\ 
\hline
\textbf{Interrogation} & Questioning or probing
for personal information
 & An interviewer asking an inappropriate question, such as marital status, during a employment interview.
 & As far as I’m the internet, not asking for private information from our customers such as addresses or any sensitive information.\\
\hline
\textbf{Aggregation} & Combining of various pieces of personal information
& A credit bureau combining an individual’s
payment history from multiple creditors.
 & Keeping unnecessary information from being exchanged at the minimum amount possible.\\
\hline
\textbf{Insecurity} & Carelessness in protecting information from leaks or improper access
& An e-commerce website allowing others to view an individual's purchase history by changing the URL (e.g. enterprivacy.com?id=123)
 & Having confidential and private information secured and stored away safely from malicious users.\\
 \hline
 \textbf{Identification} & Linking of information to a particular Individual.
 & A researcher linking medical files to the Governor of a state using only date of birth, zip code and gender.& I think it can be defined as a set of personal information of each individual that should not be accessible to other people \\
\hline
\textbf{Secondary Use} & Using personal information for a purpose other than the purpose or which is was collected 
& The U.S. Government uses census data collected for the purpose of apportioning Congressional districts to identify and intern those of Japanese descent in WWII. & Ensuring the minimum amount of data is available only to those that genuinely need it for business purposes, and that it's only available for the specified amount of time that the data is needed.\\
\hline 
\textbf{Exclusion} & Failing to let an individual know about the information that others have about them and participate in its handling or use 
& A company using customer call history, without the customer's knowledge, to shift their order in a queue (i.e. "Your call will be answer in the order [NOT] received") & to have the authority of controlling information about yourself who can or can not see. to be from from any interference, and to be able to interact with anyone I want. \\
\hline
\textbf{Breach of Confidentiality} & Breaking a promise to keep a person’s information confidential 
& A doctor revealing patient information to friends
on a social media website. & Having confidential and private information secured and stored away safely from malicious users. \\
\hline 
\textbf{Disclosure} & Revealing truthful personal information about a person that impacts the ways others judge their character or their security 
& A government agency revealing an individual’s address to a stalker, resulting in the individual’s murder. & Data must be kept safe, and users need that information to be seen only by those they authorize. \\
\hline
\textbf{Exposure} & Revealing an individual’s nudity, grief, or bodily functions 
& A store forcing a customer to remove clothing
revealing a colostomy bag. & Freedom of your own information.\\
\hline
\textbf{Increased Accessibility} & Amplifying the accessibility of personal information 
& A court making proceeding searchable on the Internet without redacting personal information. & A state where one can be sure no one else knows what they are doing \\
\hline
\textbf{Blackmail} & Threatening to disclose personal information 
& A dating service for adulters charging customers to delete their accounts. & -
\\
\hline
\textbf{Appropriation} & Using an individual’s identity to serve the aims and interests of another 
& A social media site using customer's images in advertising & Being able to be secure in your information so that none of it gets accessed or leaked by outside sources \\
\hline
\textbf{Distortion} & Disseminating false or misleading information about an individual 
& A creditor reporting a paid bill as unpaid to a credit bureau. & Privacy refers to an individual's right to control [..]on. This includes protecting sensitive data from [..], and providing individuals with the ability to access, correct, or delete their PI. \\
\hline
\textbf{Intrusion} & Disturbing an individual’s tranquility or solitude 
& An augmented reality game directing players onto private residential property. & The right to be let alone,or freedom from interference or intrusion.\\
\hline
\textbf{Decisional Inference} & Intruding into an individual’s decision regarding their private affairs 
& A payment processor declining transactions for contraceptives & The right to be let alone,or freedom from interference or intrusion.\\
\hline
\end{tabular}
\label{table:privacy_taxonomy}
\end{table*}

\end{document}